\documentclass[5p,times]{elsarticle}

\usepackage{amssymb}

\usepackage[hidelinks]{hyperref}

\usepackage{xcolor}
\usepackage{makecell}
\usepackage{graphicx}
\usepackage{subfig}
\usepackage[flushleft]{threeparttable}
\usepackage{amsmath}
\usepackage{adjustbox}
\usepackage{algorithm}
\usepackage{algpseudocode}
\usepackage{wrapfig}

\DeclareMathOperator{\erf}{erf}

\journal{Computers \& Security}

\begin{document}

\begin{frontmatter}



\title{XFedHunter: An Explainable Federated Learning Framework for Advanced Persistent Threat Detection in SDN}


\author[inst1,inst2]{Huynh Thai Thi}
\author[inst1,inst2]{Ngo Duc Hoang Son}
\author[inst1,inst2]{Phan The Duy}
\author[inst1,inst2]{Nghi Hoang Khoa}
\author[inst1,inst2]{Khoa Ngo-Khanh}
\author[inst1,inst2]{Van-Hau Pham}

\affiliation[inst1]{organization={Information Security Laboratory, University of Information Technology},
            city={Ho Chi Minh city},
            country={Vietnam}}

\affiliation[inst2]{organization={Vietnam National University},
            city={Ho Chi Minh city},
            country={Vietnam}}

\begin{abstract}
Advanced Persistent Threat (APT) attacks are highly sophisticated and employ a multitude of advanced methods and techniques to target organizations and steal sensitive and confidential information. APT attacks consist of multiple stages and have a defined strategy, utilizing new and innovative techniques and technologies developed by hackers to evade security software monitoring. To effectively protect against APTs, detecting and predicting APT indicators with an explanation from Machine Learning (ML) prediction is crucial to reveal the characteristics of attackers lurking in the network system. Meanwhile, Federated Learning (FL) has emerged as a promising approach for building intelligent applications without compromising privacy. This is particularly important in cybersecurity, where sensitive data and high-quality labeling play a critical role in constructing effective machine learning models for detecting cyber threats. Therefore, this work proposes XFedHunter, an explainable federated learning framework for APT detection in Software-Defined Networking (SDN) leveraging local cyber threat knowledge from many training collaborators. In XFedHunter, Graph Neural Network (GNN) and Deep Learning model are utilized to reveal the malicious events effectively in the large number of normal ones in the network system. The experimental results on NF-ToN-IoT and DARPA TCE3 datasets indicate that our framework can enhance the trust and accountability of ML-based systems utilized for cybersecurity purposes without privacy leakage.
\end{abstract}




\end{frontmatter}


\section{Introduction}
 Recently, the Advanced Persistent Threat (APT) is the most lethal and sophisticated threat that cyberspace must withstand. In contrast to traditional attacks, which are more opportunistic and shorter-term in nature, an APT attack is a kind of cyberattack in which attackers lurk in the intended target's network or system without being detected for a protracted period \cite{Alshamrani2019_apt_sur, stojanovic2020apt_review}. These attacks are primarily carried out by well-resourced adversaries, usually nation-states or state-sponsored groups, against specific targets using advanced tactics, techniques and procedures \cite{ahmad2019strategically_apt}. They focus on high-value targets like the government, large business entities, and many other institutions. While traditional solutions are still prevalent, contemporary cyber threats like APT attacks can bypass them easily. This scenario has prompted many cybersecurity experts to work towards developing next-generation solutions that can effectively detect and neutralize APT attacks \cite{talib2022apt_beacon_review, do2021novel_apt_DL, coulter2022domain_apt_detection, Xiong2022_Conan_apt_Detection}.

On top of state-of-the-art solutions, the Intrusion Detection System (IDS) remains a crucial component in APT detection. In the Network-based IDS (NIDS) approach, a detector with a global view can quickly identify network-wide attacks that target multiple systems. These abilities allow the NIDS to analyze network traffic for abnormal or suspicious patterns that could be an APT attack. However, identifying new patterns that signify malicious behavior has become increasingly difficult due to the growth in APT attacks. Another promising approach is the Provenance-based IDS (PIDS), especially the provenance graph approach. This approach has powerful semantic expression and correlation analysis capabilities that effectively detect APT-style multistep attacks, as evidenced by many studies \cite{9900181, gehani2021digging, kurniawan2022krystal}. Nevertheless, the complexity of the provenance graph presents a significant challenge and even worse, the hasty expansion of the organization is putting additional pressure on security engineers during the graph analysis process.

Hence, there is the place that the Artificial Intelligence (AI) comes into play. The AI solutions introduce a promising approach that will boost the IDS's performance to a new level, thanks to its capacity to manage the unfathomable accumulation of information and make decisions rapidly. As a result, the AI will become an important factor in optimizing the fight against crime and strengthening national security in the cyberspace \cite{radulov2019artificial, schmidt2021national}. 
Aside from the advantages, the AI-based IDS, especially the Deep Learning (DL)-based IDS, requires a substantial quantity of training data that must be updated regularly to maintain the performance of APT attack detector. Unfortunately, data from one organization and public sources are typically insufficient, and sharing resources among organizations would raise security and privacy concerns. Meanwhile, Federated Learning (FL), with its decentralized training capabilities without sharing local data, has emerged as an appropriate remedy to address the aforementioned problems. Any organization joining the FL training process would benefit from sharing the global model, which is aggregated by local models trained by participant parties with their private data, and then updating it with the local data of the organization. Although it is true that some studies have used the FL approach to train DL-based IDS systems, there are only a limited number of studies that have employed this approach in the context of PIDS.

Besides that, Software Defined Networking (SDN) presents a promising architecture that enables a programmable network by centralizing the network configuration process through software-based controllers rather than through manual configuration of individual devices. Moreover, the SDN controller delivers visibility into the entire network, providing a more comprehensive picture of security concerns and becoming a practical means of deploying modern cyber threat detection solutions. In fact, the potential applications of SDN in cybersecurity solutions have been shown on \cite{yurekten2021sdn,farris2018survey,zarca2020virtual}. Moreover, there have also been many studies \cite{thi2022federated,abdel2021federated,mothukuri2021federated} that have already succeeded in leveraging this aspect of SDN and FL methods to detect cyber attacks.

However, there is a problem that Deep Neural Networks (DNNs) are typically considered as ``black box" by both users and developers due to their comparative weakness in explaining their inference processes and outcomes. Nevertheless, for security analysis, the explainability and transparency of AI-based systems are particularly crucial. Understanding AI's decision-making process will improve our ability to analyze attacks, especially APT attacks. It also helps us to decrease false positive alerts and contributes to the development of more accurate IDS. Unfortunately, there are not many studies that have explored thoroughly the interpretability of IDS systems, particularly those integrated federated learning.

With the difficulties mentioned earlier in mind, we introduce \textit{XFedHunter}, an Explainable Federated Learning framework designed for APT detection in the context of SDN. Our framework features a FL-based IDS model that merges NIDS and PIDS into our APT detection system, while leveraging the SDN architecture to create reactive security measures and counter APT attacks. In our APT detection system, we use Graph Neuron Network (GNN), a category of DL methods that can handle graph data with complicated relationships and interdependencies between objects, to tackle challenges associated with provenance graph data. Moreover, we used a combination of Convolutional Neural Network (CNN) and Gated Recurrent Unit (GRU) to handle network data in the SDN environment. Then, we leverage a model-agnostic explanation \cite{ribeiro2016model} framework named SHapley Additive exPlanations (SHAP) \cite{lundberg2017unified} into the IDS system to understand AI's decision-making process. 
Beyond many related works in tackling the difficult challenge of understanding predictions from the IDS system when wrong alerts happen, we develop a methodology to determine the correctness of model predictions based on the penultimate layer's outputs of the malicious detection metrics histories. The experimental results on the NF-ToN-IoT dataset and DARPA TCE3 dataset show the efficiency of our framework to improve the interpretability of FL-based IDS against APT attacks and help the cybersecurity experts get a better understanding of IDS’s decisions. 

Our contributions in this work are summarized as follows:
\begin{itemize}
    \item Propose XFedHunter, a powerful collaborative APT hunting framework using FL in the SDN context. Our framework provides a robust NIDS system utilizing a combination of CNN and GRU (CNN-GRU) while leveraging Graph Neuron Network (GNN) into PIDS to detect APT attack patterns that are lurking in network traffic and the provenance graph.
    \item Integrate Explanable AI (XAI) into XFedHunter to analyze the prediction results to explore the factors that influence the decisions of FL-based APT attack detectors via the SHAP framework. 
    \item Design and perform a mechanism of advanced explanation analysis to determine the correctness of the model predictions based on the penultimate layer's outputs of the malicious detection metrics histories.
\end{itemize}

The remaining sections of this article are constructed as follows. Section \ref{related_work} introduced some related works in explainable AI in the context of APT detection. Next, the proposed framework and methodology are discussed in Section \ref{methodology}. Section \ref{experiments} describes the experimental settings and result analysis of APT detectors trained via the FL method and explanation methodologies applied to clarify a decision of detector. Finally, we conclude the paper in Section \ref{conclusion}.


\section{Related work} \label{related_work}

\begin{figure*}[ht]
\centering
\includegraphics[width=0.8\textwidth]{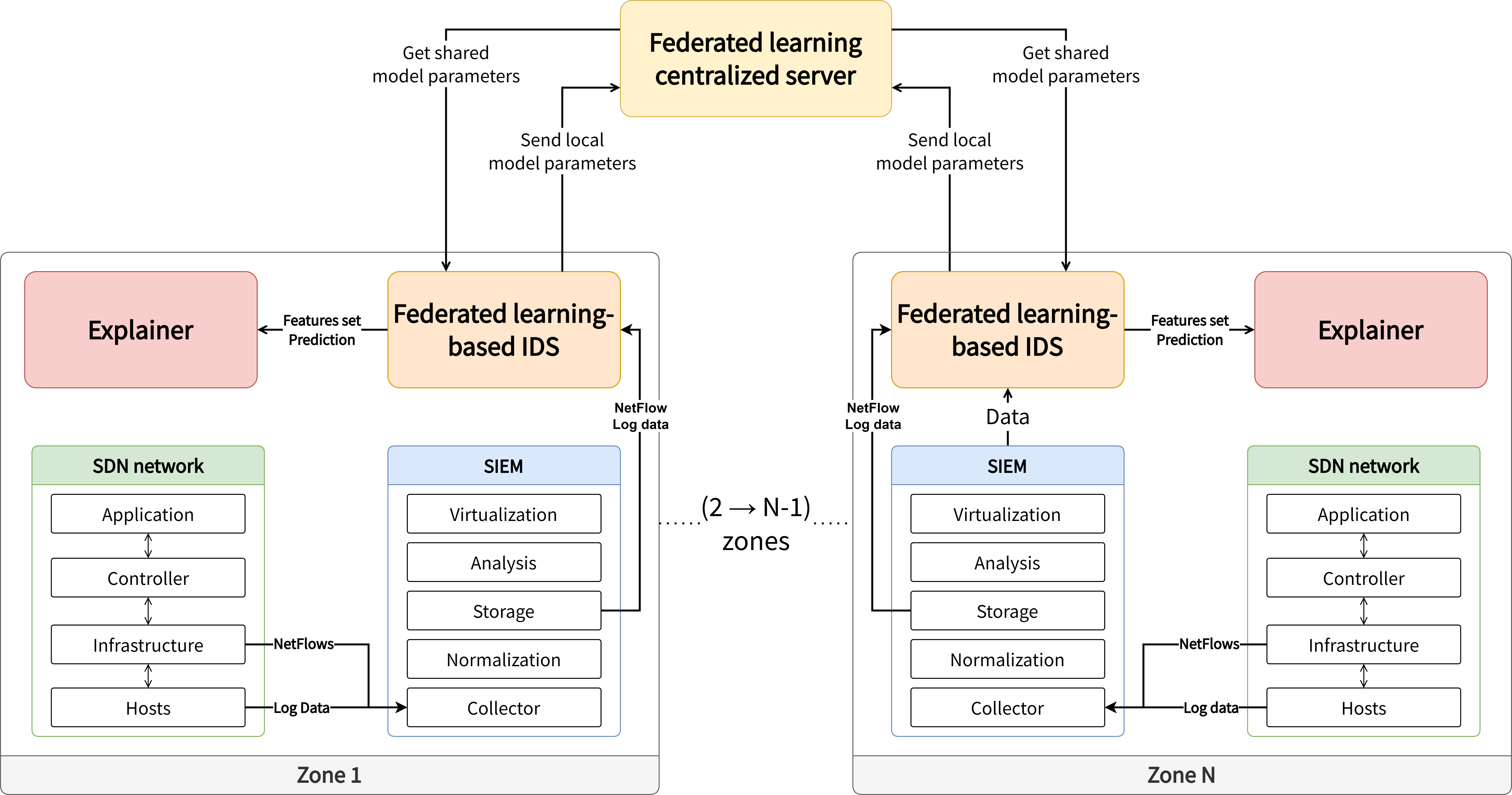}
\caption{The overall architecture of XFedHunter framework}
\label{fig:explainable-fl-sdn-model}
\end{figure*}

There are many studies that have been carried out to tackle the issues stemming from APT attacks. For instance, Yazdinejadna et al. \cite{yazdinejadna2021kangaroo} utilized an IDS module that leveraged a GRU-based algorithm. This module employed both flow-based and packet-based intrusion detection components by monitoring the packet parser and flow tables of SDN switches to detect malicious behavior effectively in SDN networks. Moreover, the authors leveraged the SDN architecture to make the IDS system ``jumps" like a kangaroo to announce the attack to other IDSs via the SDN controller, contributing to improved scalability and efficiency. In the other work, Lo and his colleagues \cite{lo2022graphsage} introduced a new method for the GNN approach called E-GraphSAGE that allows capturing both the edge features of a graph and the topological information for network intrusion detection in IoT networks using NetFlow data. The above-mentioned researches show the effectiveness of the ML-based IDS system in detecting APT attacks, but none of them leverage the FL method to enhance their detection system in the case of labeled data scarcity.

To accomplish that, Abdel-Basset et al. \cite{abdel2021federated} introduced an efficient framework that leverages a chain of DL blocks to enable cyber-threat detector capacity in detecting abnormal activity in a Microservice-based Industrial Cyber-physical System environment, while using federated learning for the training process. 
Similarly, in our previous work \cite{thi2022federated}, we had already succeeded in testing DNNs including CNN, GRU, and Long short-term memory (LSTM) to handle NetFlow data in the SDN environment while using the FL method to perform a decentralized training process. Likewise, Li and his colleagues \cite{9195012} introduced an FL-based IDS called DeepFed that utilized a combination of CNN and GRU and demonstrated to be highly effective in detecting various types of cyber threats against Industrial Cyber–Physical Systems. However, beside their success, DL-based models are becoming more and more sophisticated, and users, particularly executive staff and cybersecurity professionals at corporations, rarely interpret the model's decisions. 
As a result, the corresponding users are unable to both comprehend and verify the decisions made by DL models.

In order to handle the interpretability problems of DL-based IDS discussed earlier, Caforio et al. \cite{caforio2021leveraging} used the combination of the Grad-CAM \cite{selvaraju2017grad} explanations with the nearest-neighbor search to clarify normal and attack behavior in the network traffic and improved the accuracy of the CNN decisions. However, Grad-CAM was a model-specific explanation \cite{ribeiro2016model} method that only fits with the CNN model family. And in another work \cite{mahbooba2021explainable}, Mahbooba and his colleagues explored the interpretable side of the classification algorithm implemented for IDS using a decision tree that was considered a highly interpretable model \cite{gilpin2018explaining}. Nevertheless, similar to \cite{caforio2021leveraging} problem, the authors' method only worked with the decision tree model.

To overcome the model restriction issues, several attempts on the model-agnostic explanation method have been conducted \cite{9877919, seale2022explainable, ahmed2022artificial}. Typically, Houda and his colleagues \cite{9825734} proposed a novel DL-based IDS architecture to protect IoT-based networks against the new emerging IoT-based attacks. Then, the authors applied Explainable AI (XAI) techniques, such as SHAP, RuleFit \cite{friedman2008predictive}, and LIME \cite{ribeiro2016should}, into their proposed architecture to optimize the interpretation of decisions made by any DL-based IDS system. Finally, they validated the feasibility of the proposed framework with NSL-KDD and UNSW-NB15 datasets. Similarly, in the work of Wang et al. \cite{wang2020explainable}, an XAI framework was created utilizing the SHAP technique to increase the transparency and explainability of any IDS system. Also, the authors created two classifiers (one-vs-all and multi-class) and compared their interpretations. Lastly, the feasibility of their architecture was demonstrated by using the NSL-KDD dataset. These studies demonstrate how well DL-based IDS systems can be interpreted using model-agnostic explanation techniques, particularly the SHAP framework. Unfortunately, there is a lack of research that focuses on interpreting predictions generated by FL-based IDS systems. Furthermore, none of these studies examines the interpretation of predictions in situations in which there is a mixture of explanations for both true and false predictions, nor do they propose a methodology to address this issue.

In this research, we propose the XFedHunter framework that leverages SHAP to explain and interpret the results of FL-based IDS while enhancing its detection performance with state-of-the-art NIDS and PIDS in the context of SDN. Besides that, inspired by the research conducted by Fidel et al. \cite{9207637} in using the XAI signature created from the outputs of the penultimate layer to detect adversarial attack samples, we introduce a novelty to solve the challenge of interpreting explanations when a mixture of explanations for both true and false predictions happen. However, unlike Fidel and his colleagues, our proposed mechanism directly uses the penultimate layer's outputs of the malicious detection metrics histories instead of their proposed XAI signatures.
\section{The System Model of XFedHunter } \label{methodology}

\subsection{The architecture of APT Hunting System: XFedHunter}

Developing a reliable APT detection system is time-con\-suming and makes this research too overwhelming. To alleviate this burden and ensure that our framework can detect APTs effectively, we have streamlined the APT hunting system from our previous work \cite{thi2022federated} and incorporated it into XFedHunter, our proposed framework. In particular, our proposed APT hunting system has four components, shown as a zone in Fig. \ref{fig:explainable-fl-sdn-model}, including an SDN network, a SIEM system, a FL-based IDS model, and an explainer module.


\subsubsection{SDN network}
In our architecture, the SDN network is viewed as a key environment where harmful activity could be present in the flow data in the network traffic or log format in the host, which must be captured, collected, and analyzed. To accomplish this, for the flow data, switches in the SDN network are configured to send flow data captured from the network to the SIEM collector through the NetFlow protocol. For the log data that would be used to generate provenance graph data, we use the SIEM agent, a software installed on a host or endpoint device that collects and sends system logs to the SIEM collector of the SIEM system.

\subsubsection{SIEM system}
The data collected from the SDN network is processed and standardized to fit the needs of the FL-based IDS system, then it is stored in the SIEM data storage. This stored data can be used for visualizing and analyzing suspicious hosts, allowing organizations to quickly respond to possible APT attacks. Additionally, to prevent missing any attacks, the stored data is also transmitted to an FL-based IDS system for more advanced APT detection. 

\subsubsection{FL-based IDS model}
Beyond our previous work just using simple DNNs like CNN, GRU, and LSTM, we utilize the advanced DNNs to enhance our IDS's performance. Particularly, for the NIDS model, we use a variant of DeepFed model inspired by \cite{gonzalez2018fedids} that combines CNN and GRU to predict malicious activity in the SDN network in the NetFlow format. For the PIDS model, we leverage the E-GraphSAGE model proposed in \cite{lo2022graphsage} to handle the provenance graph data parsed from the collected system log data. The details of the used DNNs will be discussed in Section \ref{sec:experimental_setting}. Both of CNN-GRU and E-GraphSAGE are trained with the scheme of FL as mentioned in Section \ref{section_FLScheme}.

\subsubsection{Explainer module}
In this module, we leverage the SHAP framework to explain prediction decisions of IDS and our proposed method to verify the correctness of the model's results. Nevertheless, the SHAP framework needs consistent interaction with the IDS, which may significantly impede the IDS system's performance. Therefore, we recommend that the module should include a replica of the original IDS model and perform explanations on that model. The module architecture will be thoroughly discussed in Section \ref{sec:explainer_module}.

\subsection{Federated Learning Scheme for Hunting Model} \label{section_FLScheme}
Our proposed FL model, presented in Figure ~\ref{fig:explainable-fl-sdn-model}, utilizes the FedAvg algorithm \cite{mcmahan2017communication} for model aggregation to conduct decentralized training among data holders. In our model, each party engaged in the training process is considered a zone, with the FL-based IDS system serving as the primary component for communication with the centralized server during the training phase. Typically, each training cycle performs these following steps:
\begin{itemize}
    \item First, the party will interact with the centralized server for federated parameters. In the FedAvg algorithm, these parameters will be the global model’s weights or a randomly generated model’s weights.
    \item Next, the collaborating party trains the local model with their dataset based on the parameters obtained from the centralized server.
    \item After the training process is completed, the collaborator will send the local parameters, which include the local model's weights after being trained and the dataset's size using to train the local model, to the centralized server. 
    \item On the server side, after receiving all parties' local parameters, a new global model’s weights will be calculated by Eq.(\ref{eq:agg_fedavg}).
    \item Finally, the server sends the new global model’s weights as the federated parameters to collaborating parties for continuous training or use.
\end{itemize}

In the FedAvg algorithm, the aggregation function is defined as Eq.(\ref{eq:agg_fedavg}):
\begin{equation}
	w = \sum_{k=1}^{K} \frac{n_k}{n} w_k
	\label{eq:agg_fedavg}
\end{equation}
where $w$ is the new global model's weights, $K$ is the number of parties such as operational networks, cybersecurity organizations, participating in the federated training process, $w_k$ is the local model's weights for the k-th collaborator, $n_k$ is the dataset's size used to train the local model for the k-th collaborator, and $n$ is the total dataset's size for all clients.


\subsection{Explainer Module} \label{sec:explainer_module}
\begin{figure*}[!ht]
\centering
\includegraphics[width=0.8\linewidth]{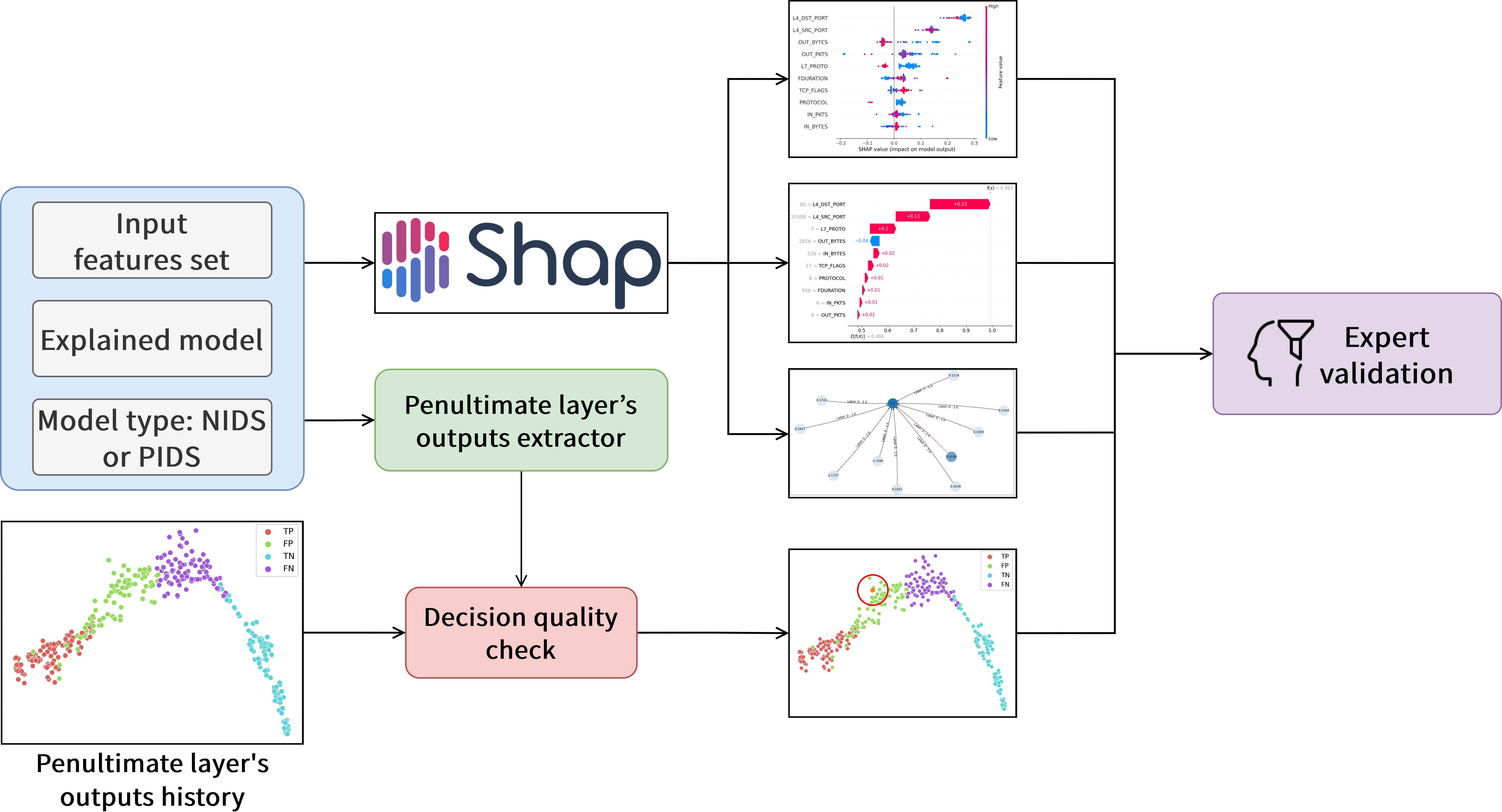}
\caption{The architecture of the Explainer module in XFedHunter framework}
\label{fig:explainer-module}
\end{figure*}

In this module, we take advantage of the SHAP framework and our proposed methodology to enhance the interpretability of the IDS system's decision, the detailed architecture of Explainer module is shown in Fig.~\ref{fig:explainer-module}.

\subsubsection{Explaining predictions with SHAP framework}
A game theoretic approach that connects optimal credit allocation with local explanations using Shapley values. In simple terms, Shapley values provide a way to fairly distribute the payoff from a cooperative game among its players \cite{winter2002shapley}. In the SHAP framework, an explanation with Shapley values for a prediction made by the model $f$ for instance $x$ having $M$ features can be defined through function $g$ as follows:
\begin{equation} \label{eq:SHAP_explain_function}
    g(z') = \phi_0 + \sum_{j=1}^{M} \phi_j z'_j
\end{equation}
where the Shapley value of $\phi_j$ represents the contribution of the j-th feature in $x$, $z'$ is a binary vector presenting a simplified set of features for $x$ ($z' \in \{0, 1\}^M$). The values in $z'$ determine which $\phi_j$ should be included in the explanation and which should not (1 for present and 0 for not present). Finally, $\phi_0$ represents the model output with all simplified inputs toggled off (i.e., all values in $z'$ are marked as 0).

In Eq.(\ref{eq:SHAP_explain_function})), the value of $\phi_i$ can be defined as follows:
\begin{equation} \label{eq:shapley_values}
    \phi_i = \sum_{z' \subseteq x' \backslash {i}} \frac{|z'|!(M - |z'| - 1)!}{M!} [f(h(z')) - f(h(z' \cup i))]
\end{equation}
where $x'$ is a simplified set of features for $x$ when all inputs represent ($x' = \{1\}^M$), $x' \backslash i$ denotes setting $x'_i=0$, $z' \cup {i}$ denotes setting $z'_i=1$, $|z'|$ is the number of non-zero entries in $z'$, and $h(z')$ is a reconstruction function that reconstructs a new set of features from the simplified set $z'$. If the i-th value in $z'$ is 1, the i-th feature value in the reconstructed data will be the same as that in the original data $x$, and if the i-th value in $z'$ is 0, the i-th feature value will be masked out and replaced with a random value or the value of the i-th feature from an instance in the dataset.

The SHAP framework also has many variant algorithms used to calculate $\phi_i$. Each of these algorithms has its own benefits and drawbacks, so the selection of the appropriate variant needs to be based on the particular scenario. In this module, we opted to use KernelSHAP for interpreting the decision made by NIDS and GradientSHAP for interpreting the decision made by PIDS.

KernelSHAP is a combination of Linear LIME and Shapley value taxonomies that provides highly accurate explanations and is well-suited to many types of DL models. However, the drawback is that KernelSHAP can be computationally intensive and necessitates background datasets to generate the explanations. In this variant, instead of using Eq.(\ref{eq:shapley_values})) to calculate Shapley values for each feature, KernelSHAP uses the LIME approach to fit Shapley values for an instance into coefficients of a local linear explanation model. To use the LIME approach to accomplish that, the loss function $L$ is redefined to update the explanation function $g$, as demonstrated in Eq.(\ref{eq:kernel_shap_loss_function})): 
\begin{equation} \label{eq:kernel_shap_loss_function}
    L(f, g, \pi_x) = \sum_{z' \in Z} [f(h(z')) - g(z')]^2 \pi_x(z')
\end{equation}

Therein, the local kernel $\pi_x(z')$ in Eq. Eq.(\ref{eq:kernel_shap_loss_function}), is defined by Eq. Eq.(\ref{eq:kernel_shap_function})):
\begin{equation} \label{eq:kernel_shap_function}
\end{equation}

After fitting the model $g$ by optimizing the loss function $L$, we will obtain all the needed Shapley values from the coefficients of $g$.

Different from KernelSHAP, GradientSHAP implements the expected gradients, a SHAP-based version of the integrated gradients \cite{sundararajan2017axiomatic}, to estimate the SHAP value of the i-th feature by computing the average gradient of the output regarding the input, weighted by the contribution of each sample to the SHAP value. Equation Eq.(\ref{eq:gradient_SHAP_shapley_values})) is used to estimate the SHAP value with the expected gradients as shown below:
\begin{equation} \label{eq:gradient_SHAP_shapley_values}
    \phi_i \approx \frac{1}{N} \sum_{j=1}^N \nabla f(x^j) \cdot w_i
\end{equation}
where $N$ is the number of samples used to estimate the SHAP value, $\nabla f(x^j)$ is the gradient of the output regarding the input for the sample $x^j$, and $w_i$ is a weight that represents the contribution of each sample to the SHAP value.

The weight $w_i$ in Eq.(\ref{eq:gradient_SHAP_shapley_values})) is calculated as Eq. Eq.(\ref{eq:gradient_SHAP_weight})):
\begin{equation} \label{eq:gradient_SHAP_weight}
    w_i = \frac{x_i - \bar{x_i}}{\sum_{k=1}^{M} (x_k - \bar{x_k})}
\end{equation}
where $x_i$ is the value for the i-th feature of $x$, and $\bar{x}_i$ is the baseline value for the i-th feature. The baseline value of a feature can be 0, the mean of all i-th feature values in the dataset, etc., depending on the specific application and the nature of the input features.

\subsubsection{Method of decision quality checking}
With the participation of the SHAP framework, we can produce an explanation for evaluating predictions generated by our APT detection system. However, our experiential explanation shown in Fig.~\ref{fig:TP-TN-FN-FP} indicates that the summary explanations for
\begin{algorithm}
\caption{The proposed decision quality checking algorithm}\label{alg:proposed-methodology}
\begin{algorithmic}[1]
\Require Instance $x$, model $f$, background data $S$
\Ensure The label for $f(x)$

\Function{SplitData}{$f, S$}
    \State $TPs, TNs , FPs, FNs \gets (), (), (), ()$
    \ForAll{$s \in S$}
        \State $y\_pred \gets f(s)$
        \If{$y\_pred$ \textbf{is} $``TP"$}
            \State $TPs.append(s)$
        \ElsIf{$y\_pred$ \textbf{is} $``TN"$}
            \State $TNs.append(s)$
        \ElsIf{$y\_pred$ \textbf{is} $``FP"$}
            \State $FPs.append(s)$
        \Else 
            \State $FNs.append(s)$
        \EndIf
    \EndFor
    \State \textbf{Return} $(TPs, TNs , FPs, FNs)$
\EndFunction
\State
\Function{GernerateTrainData}{$f, categorized\_data$}
    \State $train\_data \gets ()$ 
    \ForAll{$subset \in categorized\_data$}
        \State $penultimate\_based\_data \gets f^{-1}(subset)$ \Comment{$f^{-1}(subset)$ would return a list including the output of the penultimate layer of model $f$ for all sample $subset$.}
        \State $train\_data.append(penultimate\_based\_data)$
    \EndFor
    \State \textbf{Return} $train\_data$
\EndFunction
\State
\Function{Classifier}{$x, S'$}
    \State $...$ \Comment{classification method}
    \State \textbf{Return} $predicted\_class\_index$
\EndFunction
\State
\State $labels \gets (``TP", ``TN", ``FP", ``FN")$
\State $subsets \gets \Call{SplitData}{f, S}$
\State $train\_data \gets \Call{GernerateTrainData}{f, subsets}$
\State $x' \gets f^{-1}(x)$ \Comment{$f^{-1}(x)$ would return the output of the penultimate layer of model $f$ for the instance $x$}
\State $idx \gets \Call{Classifier}{x', train\_data}$
\State \textbf{Output:} $labels[idx]$
\end{algorithmic}
\end{algorithm}
 FN predictions are nearly the same pattern as TP predictions. Similarly, the same results can be found when comparing the summary explanations for TN predictions with FP predictions. The overlap between explanations for true and false predictions makes the security analyst confused and adds more burden to the analysis process in a real-world situation. To tackle this problem, we propose a method of checking decision quality shown in Algorithm \ref{alg:proposed-methodology} to verify the decisions made by the explained model.

In the proposed method, we first classify the input based on its prediction results into four categories in the confusion matrix \cite{stehman1997selecting}. That means the input will be classified into one of the following four categories: TP, FP, TN, and FN, as we defined in Section \ref{sec:performance_metric}. From the classification, we can remove the confusion about the explanation of wrong predictions. After splitting the dataset into four subsets based on their prediction category, we fetch all the subsets above into the IDS model and extract the penultimate layer's outputs to create a penultimate-based dataset with four classes. Finally, we will use the created dataset as a training dataset for the classifier. The reason that we decide to use the penultimate layer’s output rather than the values of the original dataset is because the neurons of this layer actually form high-level features of the original input \cite{Goodfellow-et-al-2016}, as shown in Fig. \ref{fig:compare-original-penultimate}. With this, we can easily explore the connection between its values and model prediction.

\section{Experiments and Analysis} \label{experiments}

\subsection{Dataset and Preprocessing}
\subsubsection{Dataset}

For the evaluation of our approach, we use two datasets, the first one is a Netflow-based dataset called NF-ToN-IoT  \cite{sarhan2020netflow}, and the second one is the DARPA Transparent Computing Engagement 3 (TCE3) events dataset\footnote{DARPA TCE3: \url{https://github.com/darpa-i2o/Transparent-Computing/blob/master/README-E3.md}}. 

The NF-ToN-IoT dataset is a part of the NetFlow v1 dataset collections, which includes NF-UNSW-NB15, NF-UQ-NIDS, NF-BoT-IoT, and NF-CSE-CIC-IDS2018 datasets\footnote{NetFlow dataset: \url{https://staff.itee.uq.edu.au/marius/NIDS_datasets}}. It consists of 12 NetFlow features listed in TABLE \ref{tab:flow-features}, and contains various types of attacks like DDoS, backdoor, scanning, etc. The dataset has a total of 1,379,274 records, where 1,108,995 (80.4\%) are attack flows and 270,279 (19.6\%) are benign ones.

The DARPA TCE3 is an exercise that consists of one scenario with multiple independent attackers. These attackers conducted various APT attacks against target systems which were monitored to collect log data throughout the exercise. Log data contains objects (files, unnamed pipes, NetFlow, etc.) and events that each of them associates with one or two objects. For our evaluation, we decided to select the collected CADETS FreeBSD system log data in the DARPA TCE3 dataset. The log data from CADETS FreeBSD system has a total of 13,880,763 events that contain benign data and attack data from several Nginx backdoors with Drakon APT attacks.

\subsubsection{Preprocessing}

Regarding the NF-ToN-IoT dataset, to guarantee the efficient training of the model, we remove source and destination IP addresses and perform some preprocessing steps on the raw data. We also apply two feature scaling formulas, and which formula used for each feature is mentioned in TABLE \ref{tab:flow-features}. The first one is min-max normalization applied to rescale the feature's value into the range of [0,1] as follows:
\begin{equation} \label{eq:mix&max-normalized}
    x_{normalized} = \frac{x-x_{min}}{x_{max}-x_{min}}
\end{equation}
where $x_{min}=0$, $x_{max}=256^{feature\_size}-1$ where $feature\_size$ is the corresponding feature size shown in TABLE~\ref{tab:flow-features}, and $x$ is the value of the current instance.

In the second formula, we utilize the normalization technique proposed by Raskovalov et al. \cite{raskovalov2022investigation} to optimize the flexibility of the feature transformation process on features that have large range values. The formula is defined as follows:
\begin{equation} \label{eq:erf-normalized}
    x_{normalized} = \erf(\frac{x}{k_w})
\end{equation}
where $k_w$ is the corresponding normalization coefficient shown in TABLE~\ref{tab:flow-features}, $\erf$ denotes the error function, and $x$ is the value of the current instance.

\begin{table}[ht]
    \centering
    \caption{Normalization method for each NetFlow feature}
    \label{tab:flow-features}
    \scalebox{0.85}{
    \begin{threeparttable}
    \begin{tabular}{|c|c|c|c|c|c|}
    \hline
    Feature & Feature size & \makecell{Normalization\\method} & \makecell{Normalization\\coefficient $k_w$} \\
    \hline
    PV4\_SRC\_ADDR & 4 bytes & - & -\\
    \hline
    IPV4\_DST\_ADDR & 4 bytes & - & - \\
    \hline
    PROTOCOL & 1 bytes & \eqref{eq:mix&max-normalized} & - \\
    \hline
    L4\_SRC\_PORT & 2 bytes & \eqref{eq:mix&max-normalized} & - \\
    \hline
    L4\_DST\_PORT & 2 bytes & \eqref{eq:mix&max-normalized} & - \\
    \hline
    IN\_PKTS & 4 bytes & \eqref{eq:erf-normalized} & 20 \\
    \hline
    OUT\_PKTS & 4 bytes & \eqref{eq:erf-normalized} & 20 \\
    \hline
    IN\_BYTES & 4 bytes & \eqref{eq:erf-normalized} & 900 \\
    \hline
    OUT\_BYTES & 4 bytes & \eqref{eq:erf-normalized} & 900 \\
    \hline
    TCP\_FLAGS & 1 bytes & \eqref{eq:mix&max-normalized} & - \\
    \hline
    FDURATION\tnote{*} & 4 bytes & \eqref{eq:erf-normalized} & 600 \\
    \hline
    L7\_PROTO & 2 bytes & \eqref{eq:mix&max-normalized} & - \\
    \hline
    \end{tabular}
       \begin{tablenotes}
            \item[*] \textit{FDURATION denotes FLOW\_DURATION\_MILLISECONDS.}
        \end{tablenotes}
    \end{threeparttable}}
\end{table}

For the DARPA TCE3 dataset, since only a minimal number of events in the log data are attacks, selecting all the events can cause some issues for model training. Instead, we select a subset of log data that contains most of the CADETS system's attacks. Next, we select events associated with both source and destination objects for the edge classification task in our methodology, yielding 237,721 events, of which 236,160 (99.3\%) are benign and 1,561 (0.7\%) are attacks. The graph is constructed from selected data using events as edges and objects as nodes. The features of nodes and edges are transformed into sentences, as shown in TABLE \ref{tab:sentence-pattern}, and then embedded in the output sentence using the pre-trained \emph{sentence\_transformers} model called \cite{reimers-2019-sentence-bert} \emph{all-MiniLM-L6-v2} \cite{reimers-2020-multilingual-sentence-bert}.

\begin{table}[ht]
    \centering
    \caption{Sentence patterns for each node and edge types}
    \label{tab:sentence-pattern}
    \scalebox{0.8}{
    \begin{threeparttable}
    \begin{tabular}{|c|c|c|}
    \hline
    Type & Object/event type & Sentence pattern \\
    \hline
    NODE & NET\_FLOW & \makecell{A ``net\_flow" node has \\a local address of \{\{local\_address\}\}, \\a local port of \{\{local\_port\}\}, \\a remote address of \{\{remote\_address\}\}, \\and a remote port of \{\{remote\_port\}\}.} \\
    \hline
    NODE & FILE & \makecell{A ``file" node has the subtype of \\``\{\{sub\_type\}\}".} \\
    \hline
    NODE & SUBJECT & \makecell{A ``subject" node the has subtype of \\``\{\{sub\_type\}\}".} \\
    \hline
    NODE & UNNAMED\_PIPE & - \\
    \hline
    EDGE & EXECUTE & \makecell{A ``execute" edge \\executed the ``\{\{exec\}\}" program, \\and its command line is ``\{\{cmd\_line\}\}".} \\
    \hline
    EDGE & ACCEPT & \makecell{An ``rename" edge \\accepted the connection from \{\{address\}\} \\with the port of \{\{port\}\}, \\and it executed the ``\{\{exec\}\}" program.} \\
    \hline
    EDGE & MODIFY\_PROCESS & \makecell{An ``modify\_process" edge \\executed the ``\{\{exec\}\}" program.} \\
    \hline
    EDGE & CREATE\_OBJECT & \makecell{An ``create\_object" edge \\executed the ``\{\{exec\}\}" program.} \\
    \hline
    EDGE & RENAME & \makecell{An ``rename" edge \\executed the ``\{\{exec\}\}" program.} \\
    \hline
    \end{tabular}
    \end{threeparttable}
}
\end{table}

\subsection{Performance Metrics} \label{sec:performance_metric}
\subsubsection{Detection metrics}

To accordingly evaluate the model prediction, we discussed and defined ground truth values as follows: true positive (TP) represents the number of correct predictions belonging to the attack class; true negative (TN) represents the number of correct predictions belonging to the benign class; False positive (FP) represents the number of normal labels that were misclassified as belonging to the attack class; False negative (FN) represents the number of attack labels that were misclassified as belonging to the normal class.

Therefore, we use four metrics as follows for our experiments:

\begin{itemize}
    \item \emph{Accuracy} is the ratio of correct and total predictions.
    \begin{equation} \label{eq:accuracy}
        Accuracy = \frac{TP + TN}{TP + TN + FP + FN}
    \end{equation}
    \item \emph{Precision} is the ratio of correct predictions having attack label and total predictions belong to attack class.
    \begin{equation} \label{eq:precision}
        Precision = \frac{TP}{TP + FP}
    \end{equation}
    \item \emph{Recall} is the correct predictions having attack label over the sum of correct predictions having attack label and misclassified belong to normal class.
    \begin{equation} \label{eq:recall}
        Recall = \frac{TP}{TP + FN}
    \end{equation}
    \item \emph{F1-score} is calculated by two times the product of precision and recall over the sum of precision and recall.
    \begin{equation} \label{eq:f1-score}
        F1-score = 2 \cdot \frac{Recall \cdot Precision}{Recall + Precision}
    \end{equation}
\end{itemize}

\subsubsection{Interpretability metrics}
Gilpin et al. \cite{gilpin2018explaining} suggested that an explanation could be evaluated based on their \emph{interpretability} and \emph{completeness}. Interpretability aimed to describe the internals of a system in a way that was understandable to humans and was tied to the cognition, knowledge, and biases of the user. On the other hand, completeness aims to provide an accurate description of how a system operates, but this could be challenging in the case of computer programs such as DNNs, which are not easily interpretable by humans. Due to this distinction in those objectives, it was difficult to achieve interpretability and completeness simultaneously. Therefore, in real-world situations, an explanation could obtain interpretability at the possible cost of completeness.

The Gilpin and his colleagues also suggested two evaluation methods for the explanations of deep network processing like the SHAP framework as follows: completeness compared to the original model and completeness as measured on a substitute task. However, several previous studies, including \cite{lundberg2017unified, ribeiro2016should, ancona2017towards}, have shown that the SHAP framework, especially the two variant algorithms used to explain the predictions of our IDS models, has already demonstrated its faithfulness to the original model through these evaluation methods. Therefore, in our experiment, we opted to use the human-based evaluation method that was also described in \cite{gilpin2018explaining} to investigate the interpretability of explanations generated by the SHAP framework, especially in cases where explanations are mixed for both true and false predictions from the perspective of a domain expert.


\subsection{Experimental Settings} \label{sec:experimental_setting}
We simulated the training process of the FL model with 10 clients on an Ubuntu 20.04 virtual machine with 6 core CPUs and 32 GB of RAM in 20 rounds. In our experiments, we use a CNN\&GRU model, a variant of DeepFed model, for the NIDS system and E-GraphSAGE for the PIDS system. TABLE \ref{tab:cnn-gru-model-architecture} and TABLE \ref{tab:e-graphsage-model-architecture} show the architecture for the CNN\&GRU model and E-GraphSAGE, respectively. Based on our experiments, we trained all clients with the following configuration to achieve optimal performance on both models: $Adam\ optimizer$ with $learning\_rate=0.001$, $epoch=25$, and $batch\_size=512$ for the CNN\&GRU model, and the same configuration for the E-GraphSAGE model, but with $epoch=100$ and no $batch\_size$. In the settings for both NF-ToN-IoT and the DARPA TCE3 dataset, we select 70\% of samples for the training set and the remaining 30\% for the testing set, while ensuring the ratio of malicious and benign samples to prevent bias in the evaluation process. Finally, the training dataset would be divided equally and distributed to all clients.

\begin{table}[ht]
    \centering
    \caption{The architecture of CNN\&GRU model}
    \label{tab:cnn-gru-model-architecture}
    \scalebox{0.8}{
    \begin{threeparttable}
    \begin{tabular}{|c|c|c|c|}
    \hline
    Layer (ID) & Activation & Output shape & Connected to  \\
    \hline
    Input (1) & - & (10, 1) & [] \\
    \hline
    Conv1D (2) & ReLU & (10, 32) & [(1)] \\
    \hline
    BatchNormalization (3) & - & (10, 32) & [(2)] \\
    \hline
    MaxPooling1D (4) & - & (10, 32) & [(3)] \\
    \hline
    Conv1D (5) & ReLU & (10, 32) & [(4)] \\
    \hline
    BatchNormalization (6) & - & (10, 32) & [(5)] \\
    \hline
    MaxPooling1D (7) & - & (10, 32) & [(6)] \\
    \hline
    Conv1D (8) & ReLU & (10, 32) & [(7)] \\
    \hline
    Batch Normalization (9) & - & (10, 32) & [(8)] \\
    \hline
    MaxPooling1D (10) & - & (10, 32) & [(9)] \\
    \hline
    Flatten (11) & - & (320) & [(10)] \\
    \hline
    GRU (12) & - & (3) & [(1)] \\
    \hline
    Concatenate (13) & - & (323) & [(11), (12)] \\
    \hline
    Dense (14) & - & (64) & [(13)] \\
    \hline
    Dense (15) & Sigmoid & (1) & [(14)] \\
    \hline
    \end{tabular}
    \end{threeparttable}}
\end{table}

\begin{table}[h]
    \centering
    \caption{The architecture of E-GRAPHSAGE model}
    \label{tab:e-graphsage-model-architecture}
    \scalebox{0.8}{
    \begin{threeparttable}
    \begin{tabular}{|c|c|c|c|}
    \hline
    Layer (ID) & Activation & Output shape & Connected to \\
    \hline
    Input (1) & - & \makecell{[Node: (1, 384),\\Edge: (1, 384)]} & [] \\
    \hline
    E-GRAPHSAGE (2) & ReLU & Node: (1, 128) & [(1)] \\
    \hline
    E-GRAPHSAGE (3) & ReLU & Node: (1, 384) & [(2)] \\
    \hline
    Dropout (4) & - & Node: (1, 384) & [(3)] \\
    \hline
    Dense (5) & - & Edge: (2) & [(4)] \\
    \hline
    \end{tabular}
    \end{threeparttable}}
\end{table}

\subsection{Experimental Results}
\subsubsection{Detection evaluation} 

The TABLE~\ref{tab:detection-metric-result} shows the detection metric results for the CNN\&GRU model on the NF-ToN-IoT dataset and the E-Graph\-SAGE model on the DARPA TCE3 dataset. Overall, we can see that the metric results for both models are very high, with the lowest values for the CNN\&GRU and E-GraphSAGE models being 0.9962 and 0.9449, respectively. Moreover, although the DARPA TCE3 dataset is dramatically unbalanced in favor of benign events, the E-GraphSAGE model still has a recall metric result of 0.9877, demonstrating the impressive performance of the model in searching for small malicious patterns like APT attacks among large benign events.
\begin{table}[h]
    \centering
    \caption{Detection performance of CNN\&GRU and the E-GraphSAGE model in XFedHunter}
    \label{tab:detection-metric-result}
    \scalebox{0.9}{
    \begin{tabular}{|c|c|c|c|c|}
    \hline
    & Accuracy & Precision & Recall & F1-Score \\
    \hline
    CNN\&GRU & 0.9969 & 0.9962 & 0.9976 & 0.9969 \\
    \hline
    E-GraphSAGE & 0.9995 & 0.9449 & 0.9877 & 0.9658 \\
    \hline
    \end{tabular}
    }
\end{table}

\subsubsection{Interpretability evaluation for CNN\&GRU model's predictions} \label{sec:interpretability-evaluation-CNN-GRU}
\begin{figure}[h]
\centering
\includegraphics[width=0.80\linewidth]{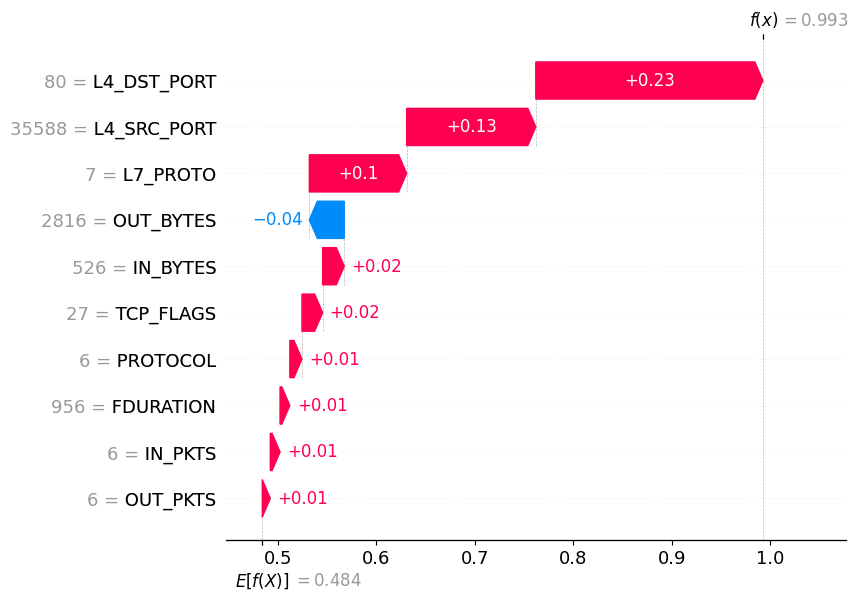}
\caption{Explanation for a prediction produced by CNN\&GRU model} 
\label{fig:SHAP-waterfall}
\end{figure}

The SHAP framework provides many visualizing functions to interpret model's decisions with Shapley values. We can explain a single prediction via the waterfall function\footnote{\url{https://shap.readthedocs.io/en/latest/example_notebooks/api_examples/plots/waterfall.html}}, as shown in Fig. \ref{fig:SHAP-waterfall}. The bottom axis of the waterfall plot describes the predicted value of the model, while each row represents how the positive (red) or negative (blue) contribution of each feature moves the model's output from the based value (the value of $E[f(x)]$ which equal $\phi_0$ in Eq.~\eqref{eq:SHAP_explain_function}) to the final output (the value of $f(x)$) based on the background dataset. 

It is apparent from Fig. \ref{fig:SHAP-waterfall} that the model classifies the predicted flow as malicious flow with a nearly perfect score of 0.993 and the L4\_DST\_PORT feature has a dominant effect (0.23) on the output. The top three most influential features following the L4\_DST\_PORT feature have positive impacts in constituting this malicious decision, while the OUT\_BYTES feature has a negative impact but does not affect the prediction too much, and the other features have no or very little impact on the decision of the model. Even though the model has made a highly confident prediction suggesting that the flow is malicious, the explanation reveals that this prediction is heavily dependent on the features L4\_DST\_PORT and L4\_SRC\_PORT. However, based on our knowledge, these features hold little significance in detecting cyberattacks in real-world scenarios. This indicates that the model might have already been fitted to the dataset context and is strongly associating malicious flows with the values of the L4\_DST\_PORT and L4\_SRC\_PORT features.

\begin{figure*}[h]%
 \centering
 \subfloat[Explanations for TP predictions]{\includegraphics[width=0.45\linewidth]{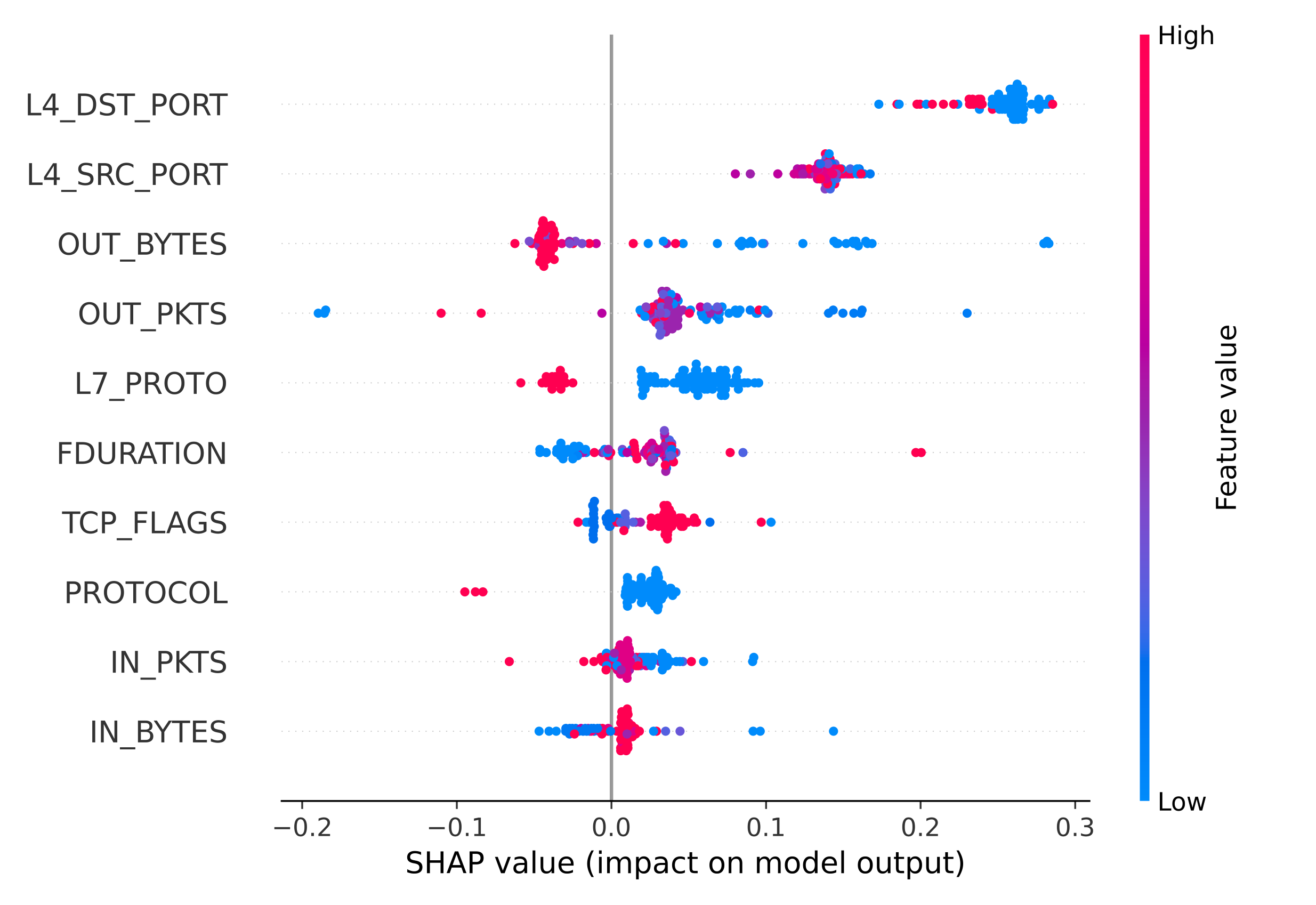}\label{fig:TP}}%
 \subfloat[Explanations for TN predictions]{\includegraphics[width=0.45\linewidth]{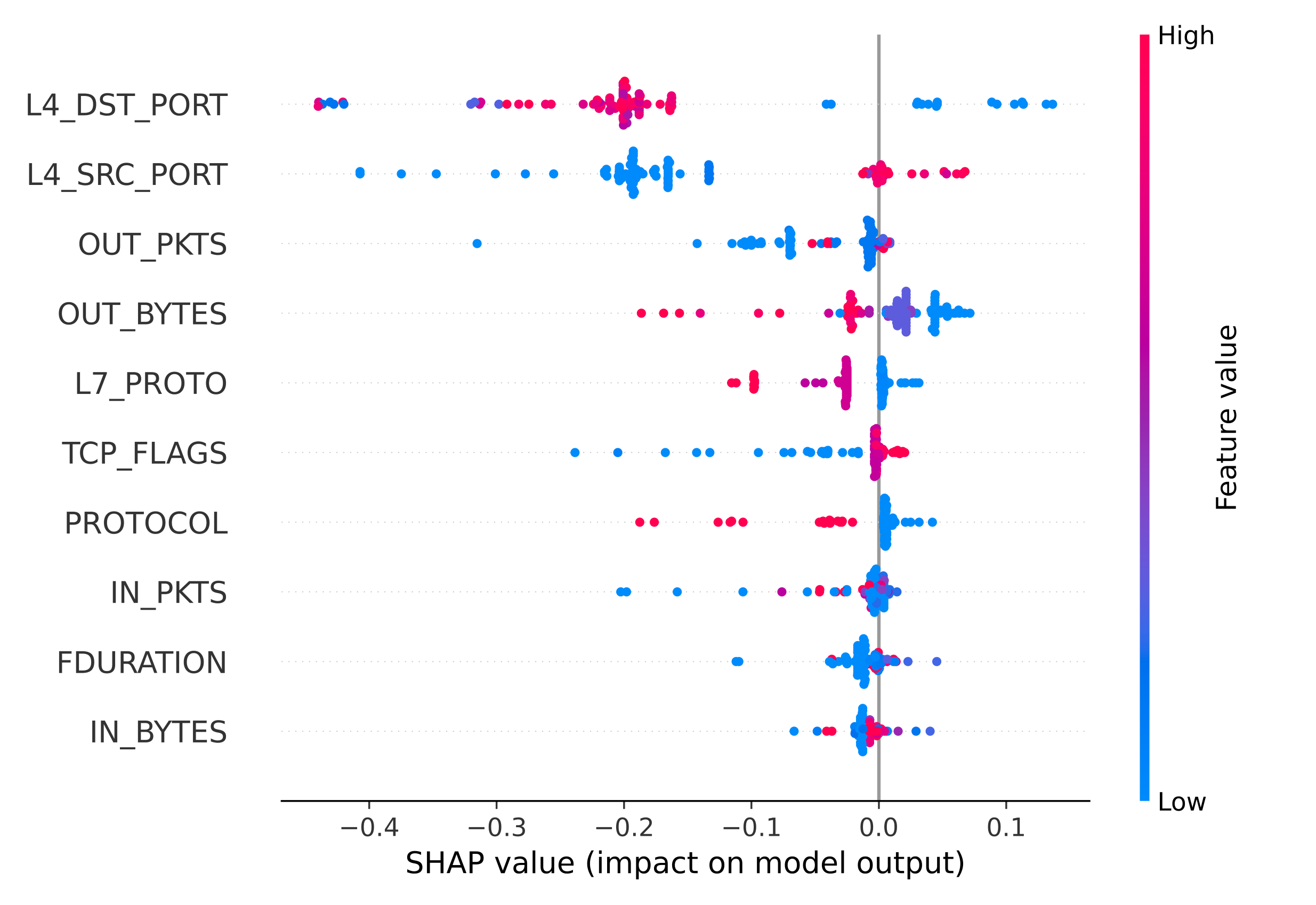}\label{fig:TN}}\\
 \subfloat[Explanations for FN predictions]{\includegraphics[width=0.45\linewidth]{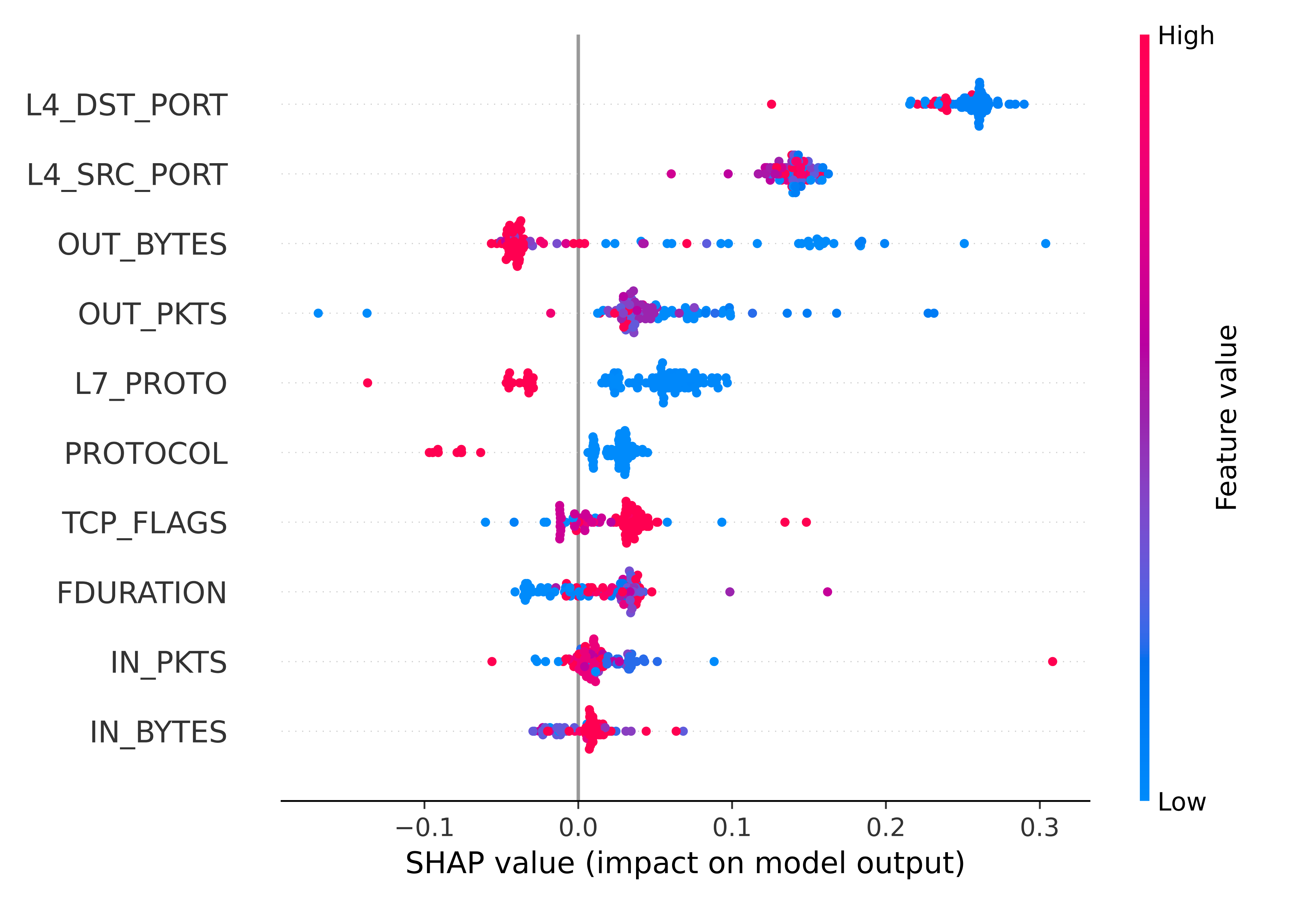}\label{fig:FN}}%
 \subfloat[Explanations for FP predictions]{\includegraphics[width=0.45\linewidth]{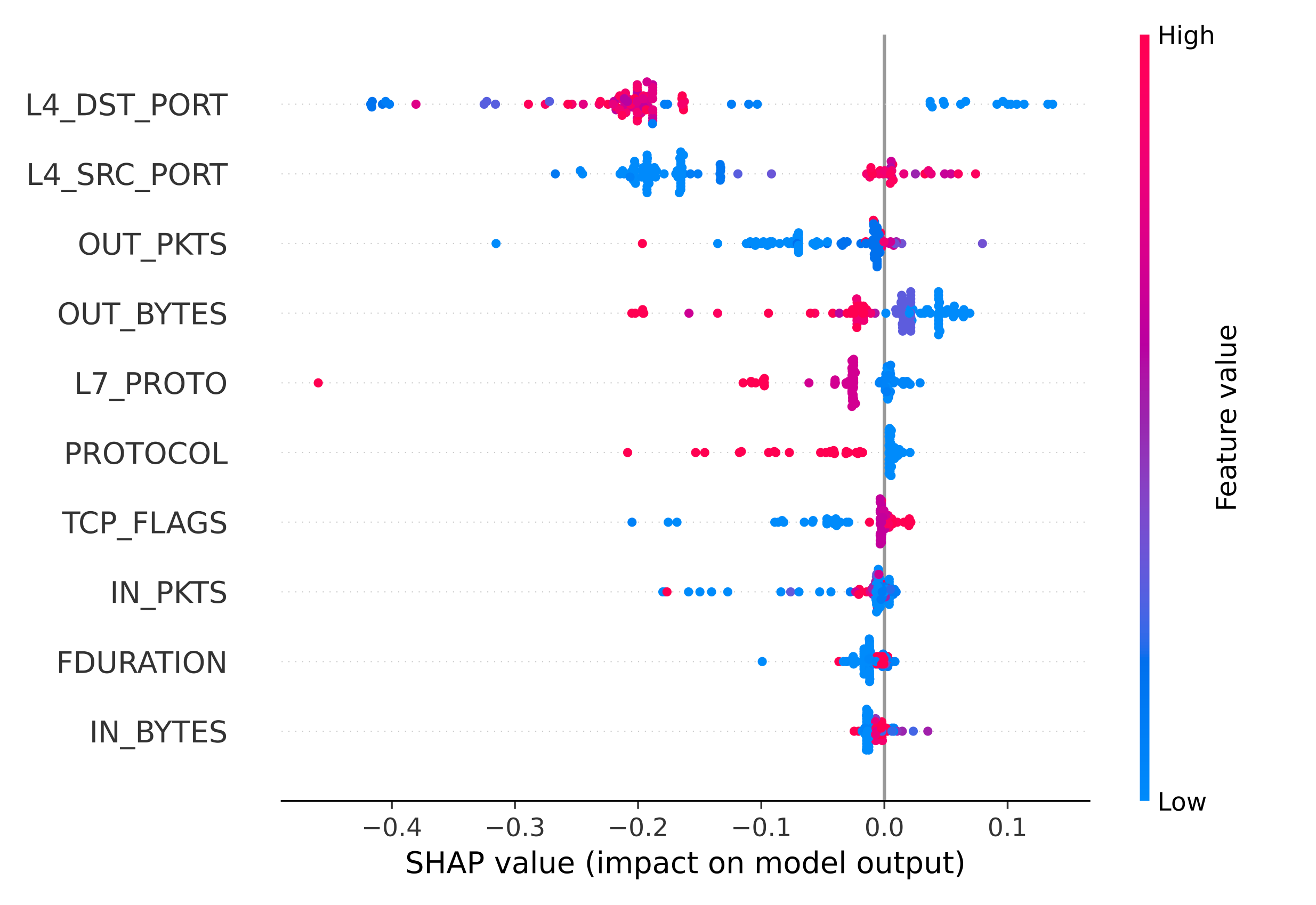}\label{fig:FP}}%
 \caption{Summary explanations for predictions produced by CNN\&GRU model based on its class (TP, TN, FN, FP)}%
 \label{fig:TP-TN-FN-FP}%
\end{figure*}

Moving forward, we use the beeswarm function\footnote{\url{https://shap.readthedocs.io/en/latest/example_notebooks/api_examples/plots/beeswarm.html}} to summarize the explanation of many predictions. So that we can understand how the top features in a dataset impact the model’s output over multiple explanations. Moreover, rather than clarifying the random model's output like other related work, we create the explanation for 4 prediction subsets based on categorical metrics (TP, FP, TN, FN). Particularly, we only extract those subsets based on predictions from CNN\&GRU model on the test data and each subset includes 100 samples. After that, we fetch those subsets into the beeswarm function and create the explanation shown in Fig. \ref{fig:TP-TN-FN-FP}. The figure displays a scatter plot where each dot represents a Shapley value for a specific feature of an instance. The features are described on the y-axis and ordered according to their average Shapley values from high to low. The x-axis indicates the Shapley value for each dot and the color reflects the corresponding value of the feature from low (blue color) to high (red color). Furthermore, the transfer from the blue to the red color demonstrates the increase in the value of the features and overlapping points jittered in the y-axis direction represent the distribution of the Shapley values per feature.


Fig.~\ref{fig:TP} and Fig.~\ref{fig:TN} show summary explanations for 100 TP predictions and 100 TN predictions generated by CNN\&GRU model, respectively. Both plots give nearly the same result for the top five most influential features, but they also have some major differences. The results shown in Fig.~\ref{fig:TP} indicate that the most influential feature (L4\_DST\_PORT) has a consistently positive impact on the model's prediction with contributed values ranging from 0.15 to 0.3 and the actual value of this feature in almost all samples is in the low range. Meanwhile, the contribution of the L4\_DST\_PORT feature in Fig.~\ref{fig:TN} ranges from -0.4 to 0.1. Notably, almost all samples exhibiting high values for this feature demonstrate a stable contribution between -0.3 and -0.2. This indicates that values in this range have a considerably negative impact on the prediction of the model and make the sample more likely to be benign. However, the contribution values are stretched out by some samples having low values for this feature that maybe indicate an uncertain model decision or a decision fitted with a certain port (the cases where the contribution has a dominant impact of roughly -0.4). 
Based on the preceding analysis, we can see that the contribution of the L4\_DST\_PORT feature for the TP predictions always has a positive impact and converge in a certain range, while the contribution of the L4\_DST\_PORT feature for the TN predictions also has the converging range but has some unexpected values affected by the sample with low values in these features.

Similarly, in Fig.~\ref{fig:TP}, the contributions for the L4\_SRC\_PORT feature in 100 TP predictions have the same characteristics as the L4\_DST\_PORT feature but converge in the lower contribution range and have the value of this feature distributed in a wider range. Besides that, the contribution of the L4\_SRC\_PORT feature in Fig.~\ref{fig:TP} also has similarities with the contribution of its most important feature, but the converging contributed point is a little lower and has the opposite feature's values compared to the most important one. The following two features in Fig.~\ref{fig:TP} and Fig.~\ref{fig:TN} are the only two features that have different positions in the top five most influential features. Moreover, a relationship between them becomes apparent when comparing these features side by side. Particularly, the OUT\_BYTES feature has a negative impact on the prediction for samples with high values in this feature and a positive impact on samples with low values. However, for TP predictions, the contributions of OUT\_BYTES heavily converge with the samples having high values of this feature, whereas for TN predictions, the contributions converge with samples having low and medium values. Furthermore, although TP and TN samples have low values in the OUT\_PKT feature, the contributions for this feature have contrary effects in each case: a positive impact for TP predictions and a negative impact for TN predictions. Similarly, we can analyze the remaining features to explore the relationship between TP and TN predictions and provide insight into how each feature affects the model's decision.

However, there are similar patterns when comparing the summary explanations for FN predictions and FP predictions shown in Fig.~\ref{fig:FN} and Fig.~\ref{fig:FP} with those for TP predictions and TN predictions. This indicates that malicious samples having similar data distributions to benign samples make our model confused and classify these samples as benign. The same conclusion can also be drawn for benign samples. From our knowledge, these confusions are understandable when put into a real-world scenario. For DoS attack instances, which we also know exist in our dataset, the DoS attack occurs when there is an enormous network flow overrun in our network. Therefore, the attack is concluded as happening or not depending on the amount of flow data and particular system. However, as in almost all research out there, we classify a model prediction as malicious if it is above a defined threshold (usually 0.5), which may not be the correct measurement threshold in all cases. Besides the above reasons, those confusions may also indicate a mislabeling in the experiment dataset or even be a signal for an adversarial attack.

In summary, the analysis conducted above enables us to comprehend the features that have the most significant contribution to the model's decision-making process through the single explanation provided by the SHAP framework. Furthermore, the summary explanations generated not only revealed the relationship between each feature's contribution and the mean of feature values, but also highlighted the similarities and contrasts between the contributions for each feature depicted in those plots. Moreover, we can see from the above analysis that although the model gives excellent performance, as evidenced by the result in TABLE \ref{tab:detection-metric-result}, the model's decision mainly relies on the two most important features, L4\_DST\_PORT and L4\_SRC\_PORT, which may be inappropriate in a real-world context. These settings can be easily evaded, especially in an APT attack where the stealth is highly valued.

\subsubsection{Interpretability evaluation for E-GraphSAGE model's predictions} \label{sec:interpretability-evaluation-E-GraphSAGE}
\begin{figure*}[h]
\centering
\includegraphics[width=0.75\linewidth]{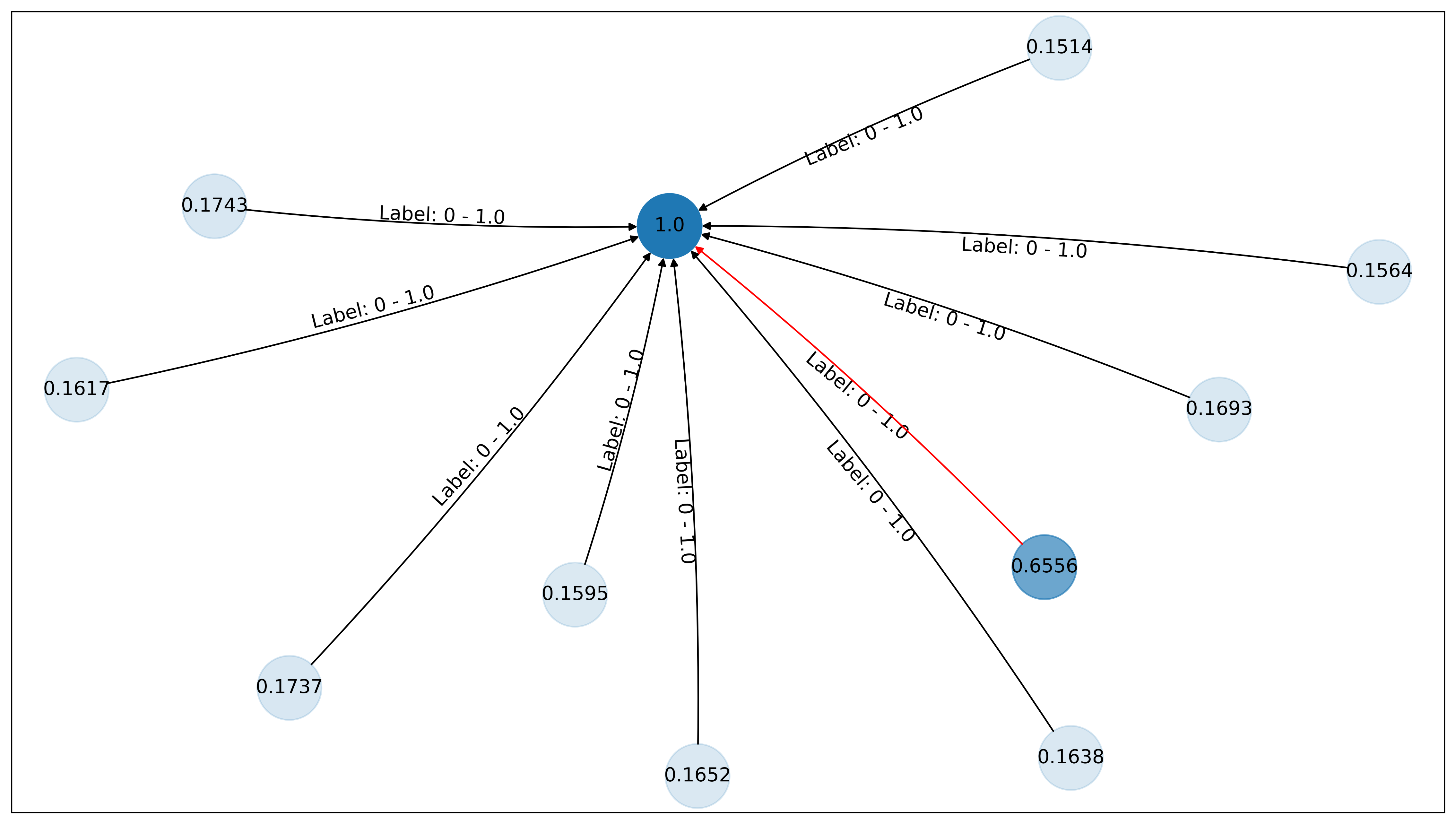}
\caption{The explanation for an edge prediction produced by E-GraphSAGE model}
\label{fig:GradientShap-E-GraphSAGE}
\end{figure*}

There are differences in the explanation for a prediction of the GNN model, specifically our E-GraphSAGE model, when compared with the CNN\&GRU model discussed earlier. Instead of focusing on important features of the predicted edge, the modern graph explanation method focuses on determining which node and which edge are more important than the others. This methodology is based on the knowledge that the prediction of an edge would depend on the attributes of the nodes and edges surrounding it, and farther nodes and edges would have less impact than those having a closer distance. Because of this taxonomy, rather than computing the importance score for each node and edge, a more effective explanation is achieved by determining the importance score of the nodes and edges in the sub-graph surrounding the predicted edge. By centering the node that has the highest importance score next to the predicted edge (the source node or the destination node), we can generate a node-centered sub-graph explanation, as shown in Fig.~\ref{fig:GradientShap-E-GraphSAGE}. The figure provides the explanation for a benign prediction edge (red color edge). Each node is labeled with its important score and the score is calculated by the sum of all feature contributions measured by GradientSHAP of that node and is scaled into [0, 1] by dividing with the highest importance score of the node in the sub-graph. Furthermore, the transformation from light blue to dark blue in the nodes of the sub-graph indicates the increase in the value of the features. Each edge is labeled with its actual class (0 is benign and 1 is malicious) and its important score is calculated similarly to the important score of a node.

It is apparent from the Fig.~\ref{fig:GradientShap-E-GraphSAGE} that the center node has a dominant effect in this sub-graph. Moreover, it is worth noting that not only our predicted edge but also the other edges that are around the center node are labeled as benign and have the same important score of 1.0, indicating the center node has a negative impact on predictions for edges around it. To better understand the explanation, we remap node and edge features to their original values. The original value for the center node is \emph{A ``file" node has the subtype of ``dir"} and the other nodes are \emph{A ``subject" node has the subtype of ``process"} and \emph{An ``modify\_process" edge executed the ``imapd" program} is the original value for all edges in the sub-graph. From the explanation and reversed values, we can conclude that the imapd program is decisively classified as benign by our models. In fact, in our experiment dataset, any events associated with the imapd program are labeled as benign events, so we can rely on the generated explanation to interpret the prediction of E-GraphSAGE model.

\subsubsection{The evaluation for the decision quality checking method}
\begin{figure*}[h]
\centering
 \subfloat[]{\includegraphics[width=0.40\linewidth]{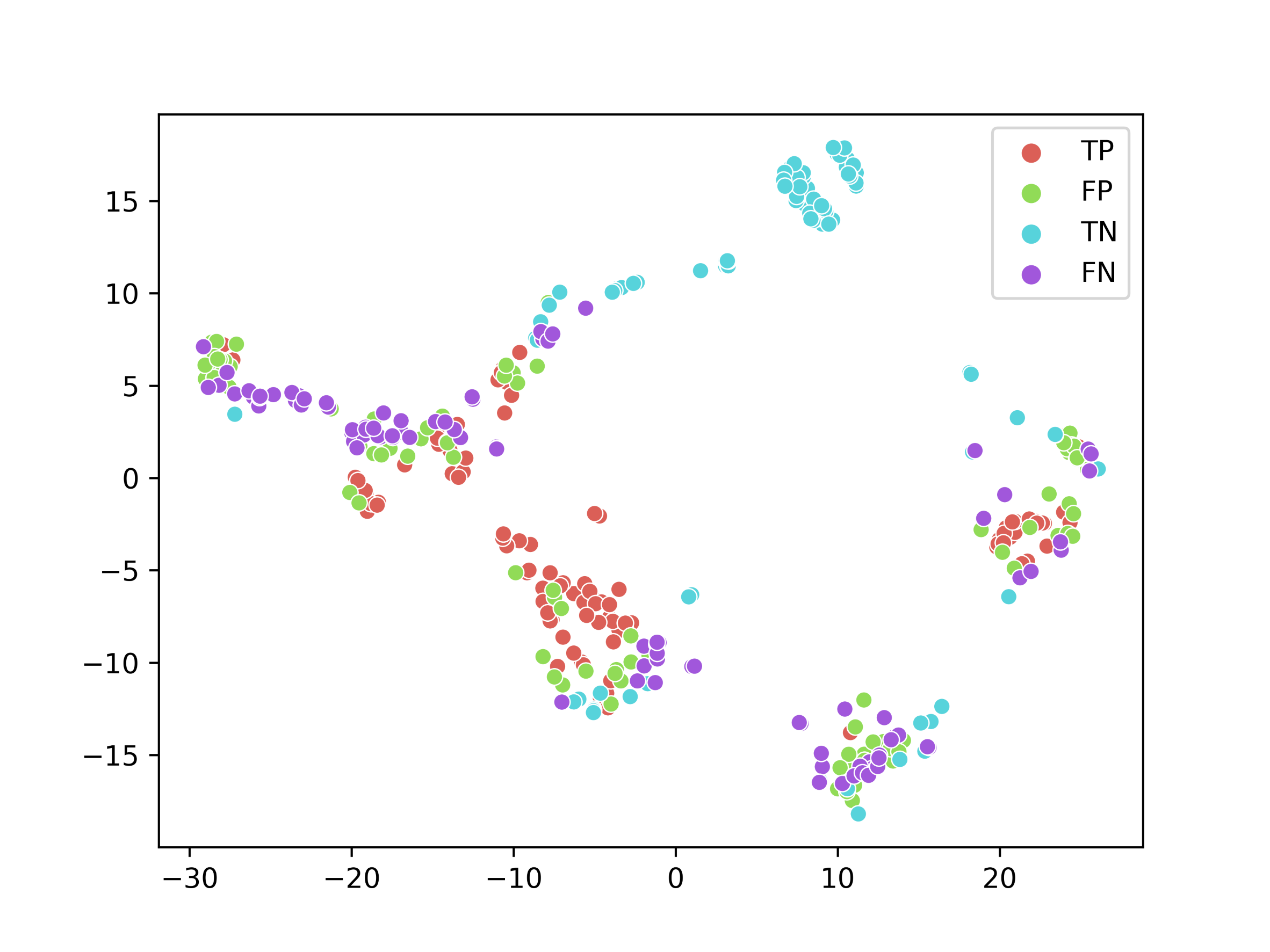}\label{fig:rawdata-tsne}}%
 \subfloat[]{\includegraphics[width=0.40\linewidth]{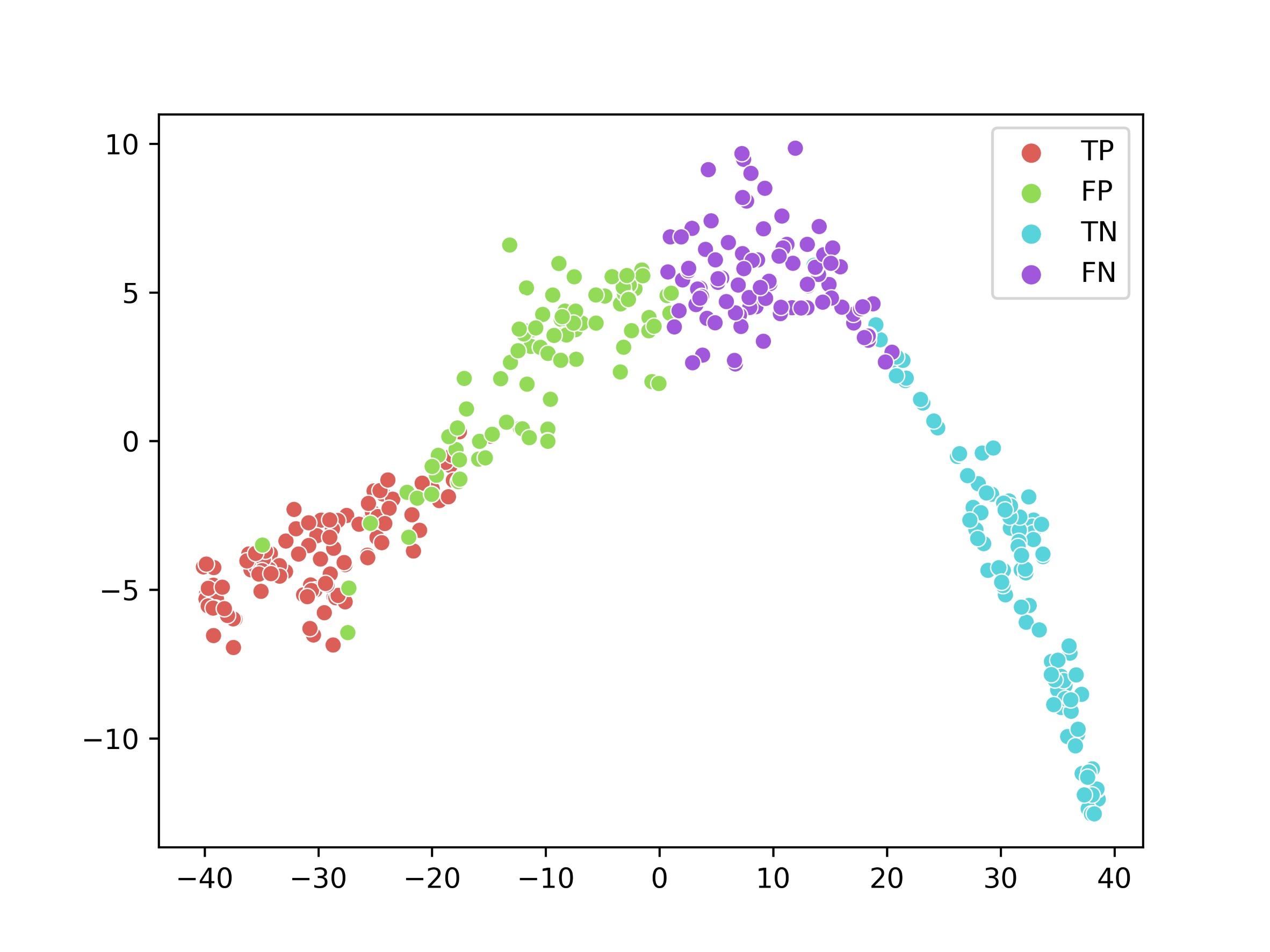}\label{fig:nep-tsne}}%
\caption{t-SNE virtualization for (a) original dataset and (b) penultimate-based dataset}
\label{fig:compare-original-penultimate}
\end{figure*}

From the testing datasets used to evaluate GRU\&CNN model in Section \ref{sec:interpretability-evaluation-CNN-GRU} and E-GraphSAGE model in Section \ref{sec:interpretability-evaluation-E-GraphSAGE}, we first create a penultimate-based dataset with the size of 400 divided equally into four classes (TP, TN, FP, FN), as outlined in Algorithm~\ref{alg:proposed-methodology}. We also use a technique called t-SNE \cite{van2008visualizing} to visualize and compare the original dataset with the penultimate-based dataset shown in Fig.~\ref{fig:compare-original-penultimate}. This technique is used here to prove that the output values of the penultimate layer actually form high-level features of the original input and are easier to use in classifying the categories of the prediction.

It is apparent from the Fig.~\ref{fig:compare-original-penultimate} that each category of the penul\-timate-based dataset is grouped better in the original dataset, although there are some penultimate-based data points located in the wrong class. This evidence implies that checking the prediction quality on the output of the penultimate layer will give a higher performance than the original data.

For the classifier that is used to measure the reliability of the prediction based on the penultimate-based dataset, we simply compute the average distance between the examined instance and instances in each class in the penultimate-based dataset. Then, the examined prediction is assigned to the class with the smallest average distance. We opt to use the average distance calculation method in our classifier because of the lack of samples for the FP and FN classes in our test dataset, although there are many samples for the TP and TN classes. Particularly for the GRU\&CNN model, we only have 300 samples for the FP class and 183 samples for the FN class in our test data. To maintain class balance, we decided to create a training dataset consisting of 400 samples and a test dataset consisting of 360 samples, with the samples in these datasets equally divided among each class. The leak of FP and FN samples is even worse for the E-GraphSAGE model, so we just have a training dataset with a size of 160 samples (50 TP samples, 50 FP samples, 50 TN samples, and 10 FN samples) and a test dataset consisting of 96 samples (30 TP samples, 30 FP samples, 30 TN samples, and 6 FN samples). Due to the small size of the created datasets, we should not use too sophisticated methods, causing unnecessary computation and decreasing the classification performance. Despite the aforementioned limitations, our approach still achieves an impressive accuracy of approximately 0.9188 in the case of the CNN\&GRU model, but the accuracy in the case of the E-GraphSAGE model is lower at 0.8557.

\section{Conclusion} \label{conclusion}
In conclusion, this paper proposes a novel explainable federated learning framework for Advanced Persistent Threat (APT) detection in Software-Defined Networking (SDN). The proposed framework, called XFedHunter, leverages local cyber threat knowledge from many training collaborators to detect and predict APT indicators and provide explanations for the ML predictions. Our framework can help cybersecurity domain experts verify the correctness of predictions from APT detectors with lesser burden through SHAP framework and a mechanism of checking model decisions on false negative and false positive cases. The experimental results on the NF-ToN-IoT and DARPA TCE3 datasets demonstrate the effectiveness of the proposed framework in enhancing the trust and accountability of ML-based systems utilized for cybersecurity purposes without privacy leakage.

The proposed XFedHunter framework addresses the challenges of APT attacks by providing explainable ML predictions to reveal the characteristics of attackers lurking in the network system. The framework also leverages the advantages of federated learning to enable effective collaborative learning while preserving data privacy. Overall, our work provides valuable insights and a practical solution for building robust and secure ML-based systems for APT detection in SDN environments. Our findings can inform future research and development of explainable federated learning frameworks for cybersecurity applications. In the future, we intend to investigate the feasibility of XAI to defend against certain adversarial attacks on deep neural networks in the context of cyberattack and malware detection.

\section*{Acknowledgment}



This research was supported by The VNUHCM-University of Information Technology's Scientific Research Support Fund.

 \bibliographystyle{elsarticle-num} 
 \bibliography{cas-refs}

   \subsection*{  } 
    \setlength\intextsep{0pt} 
  \begin{wrapfigure}{l}{21mm} 
    \includegraphics[width=1in,height=1.25in,clip,keepaspectratio]{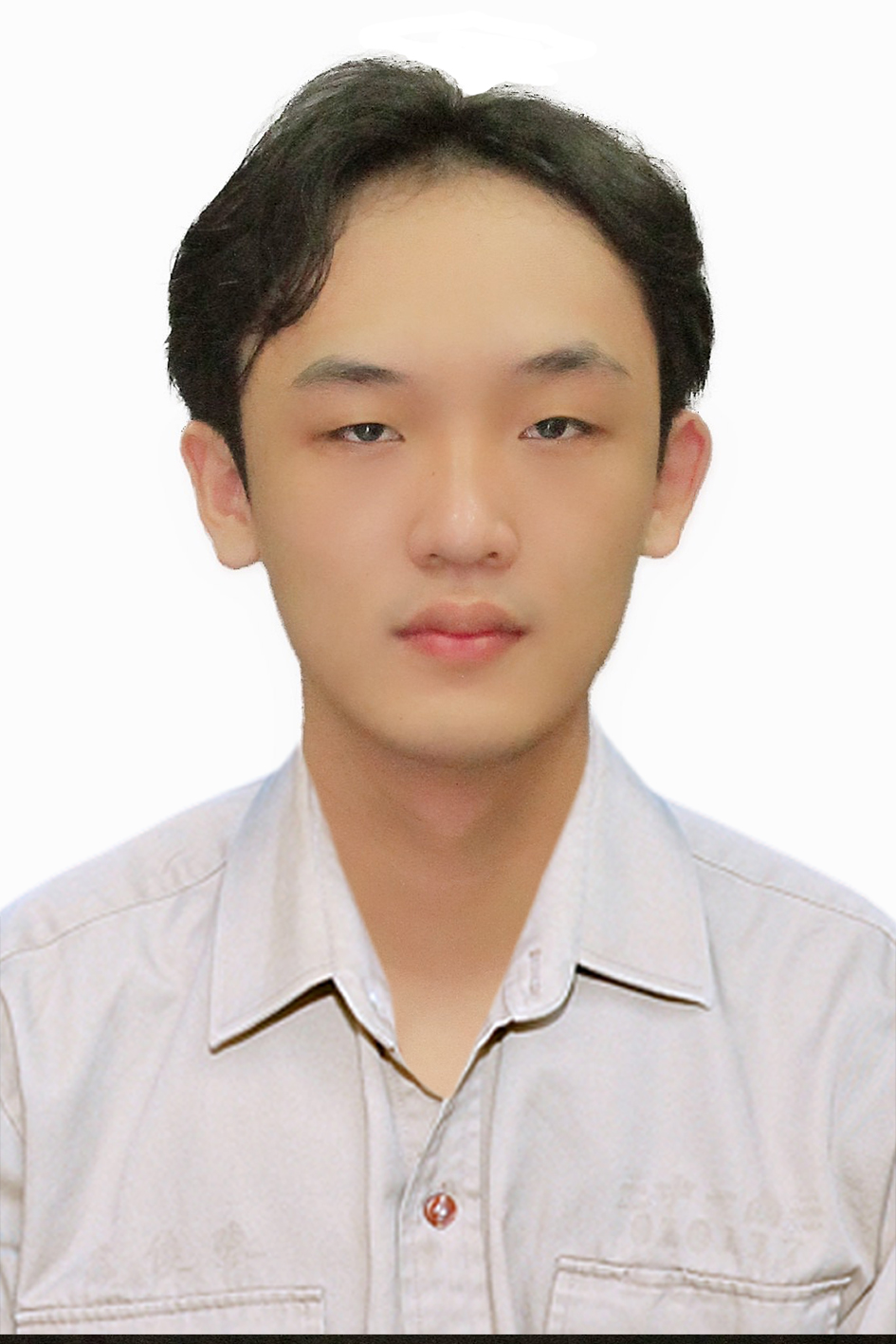}
  \end{wrapfigure}\par
  \noindent\textbf{Huynh Thai Thi} is pursuing the B.Eng. in Information Security from the University of Information Technology (UIT) Vietnam National University Ho Chi Minh City (VNU-HCM). He also works as a collaborator in Information Security Laboratory (InSecLab) UIT-VNU-HCM. His main research interests include Software Defined Networking, Explainable Artificial Intelligence and Machine Learning-based Cybersecurity.

   \subsection*{  } 
    \setlength\intextsep{0pt} 
 \begin{wrapfigure}{l}{21mm} 
    \includegraphics[width=1in,height=1.25in,clip,keepaspectratio]{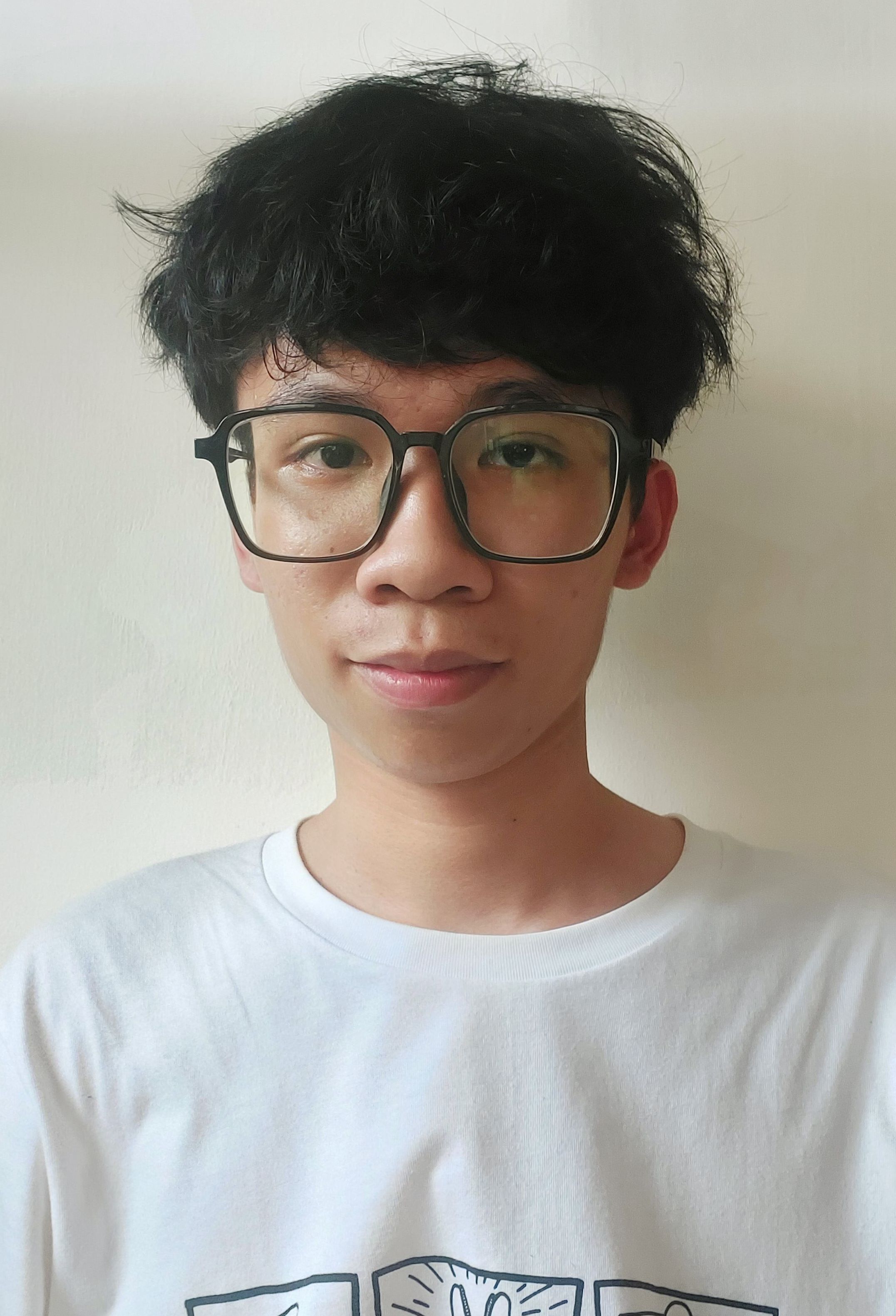}
  \end{wrapfigure}\par
\noindent\textbf{Ngo Duc Hoang Son} is pursuing the B.Eng. in Information Security from the University of Information Technology (UIT) Vietnam National University Ho Chi Minh City (VNU-HCM). He also works as a researcher intern in Information Security Laboratory (InSecLab) UIT-VNU-HCM. His main research interests include Software Defined Networking, Malware Analysis and Machine Learning-based Cybersecurity.

   \subsection*{  } 
    \setlength\intextsep{0pt} 
 \begin{wrapfigure}{l}{21mm} 
    \includegraphics[width=1in,height=1.25in,clip,keepaspectratio]{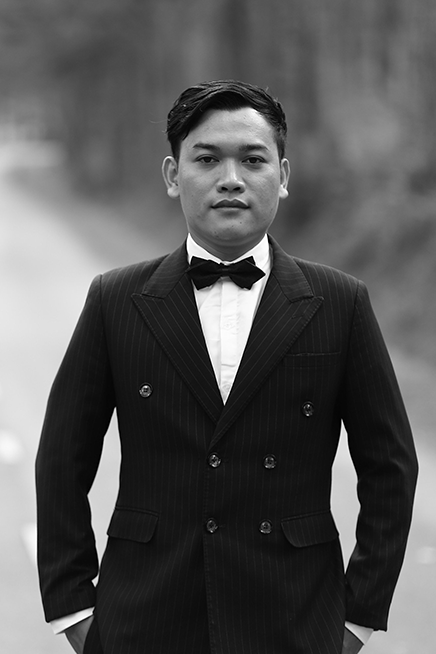}
  \end{wrapfigure}\par
\noindent\textbf{Khoa Ngo-Khanh} received the B.Eng. degree in Information Security from the University of Information Technology, Vietnam National University Ho Chi Minh City (UIT-VNU-HCM) in 2018. He is pursuing an M.Sc. degree in Information Technology at UIT-VNU-HCM. From 2022 until now, he works as a member of a research group at the Information Security Laboratory (InSecLab) at UIT. His main research interests are Information security, including Software Vulnerability, Fuzzing, Web Security.

   \subsection*{  } 
    \setlength\intextsep{0pt} 
 \begin{wrapfigure}{l}{21mm} 
    \includegraphics[width=1in,height=1.25in,clip,keepaspectratio]{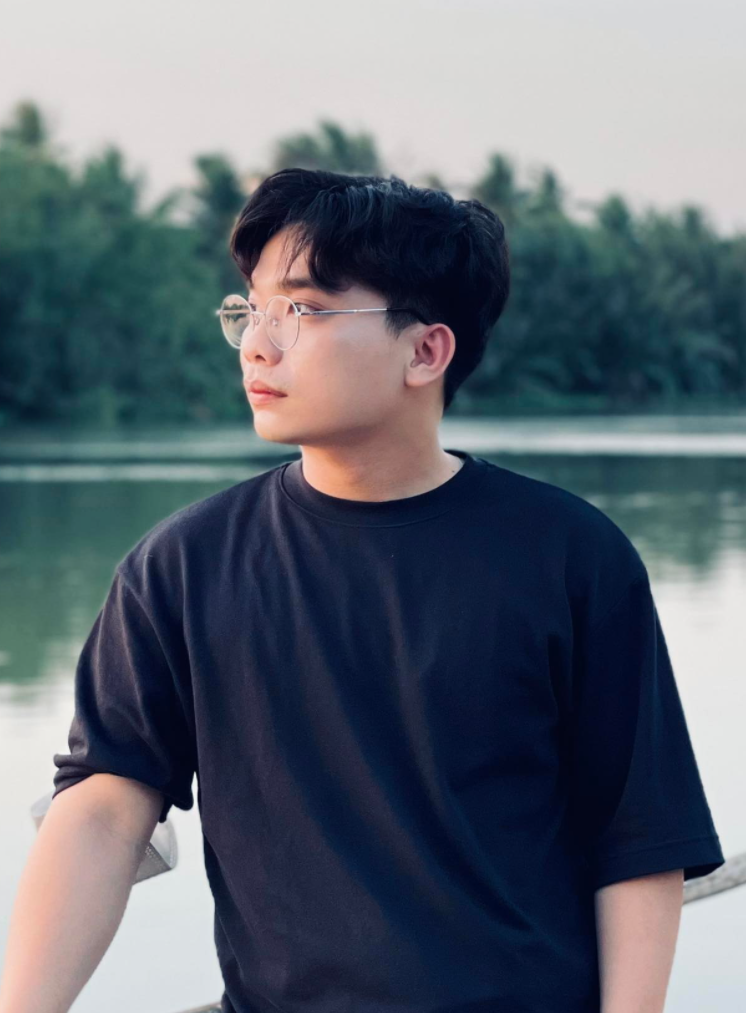}
  \end{wrapfigure}\par
\noindent\textbf{Nghi Hoang Khoa}
received the B.Eng. degree in Information Security from the University of Information Technology, Vietnam National University Ho Chi Minh City (UIT-VNU-HCM) in 2017. He also received an M.Sc. degree in Information Technology in 2022. From 2018 until now, he works as a member of a research group at the Information Security Laboratory (InSecLab) at UIT. His main research interests are Information security, malware analysis, Android security, and Intrusion Detection System.

   \subsection*{  } 
    \setlength\intextsep{0pt} 
 \begin{wrapfigure}{l}{21mm} 
    \includegraphics[width=1in,height=1.25in,clip,keepaspectratio]{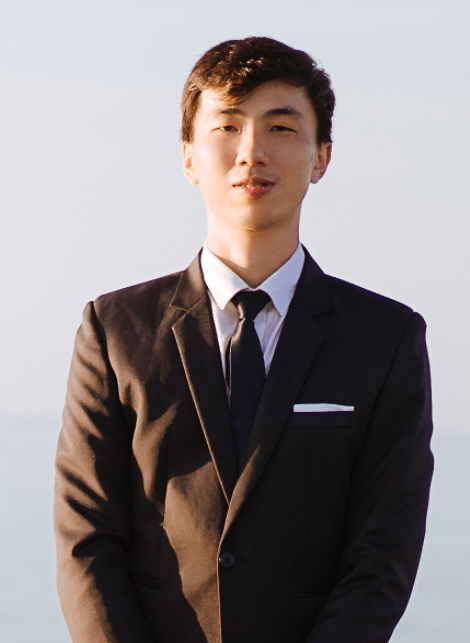}
  \end{wrapfigure}\par
\noindent\textbf{Phan The Duy}
received the B.Eng. and M.Sc. degrees in Software Engineering and Information Technology from the University of Information Technology (UIT), Vietnam National University Ho Chi Minh City (VNU-HCM) in 2013 and 2016 respectively. Currently, he is pursuing a Ph.D. degree majoring in Information Technology, specialized in Cybersecurity at UIT, Hochiminh City, Vietnam. He also works as a researcher member in Information Security Laboratory (InSecLab), UIT-VNU-HCM after 5 years in the industry, where he devised several security-enhanced and large-scale teleconference systems. His research interests include Information Security \& Privacy, Software-Defined Networking, Malware Detection, Digital Forensics, Adversarial Machine Learning, Private Machine Learning, Machine Learning-based Cybersecurity, and Blockchain.

   \subsection*{  } 
    \setlength\intextsep{0pt} 
 \begin{wrapfigure}{l}{21mm} 
    \includegraphics[width=1in,height=1.25in,clip,keepaspectratio]{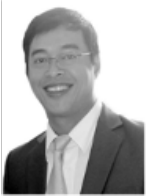}
  \end{wrapfigure}\par
\noindent\textbf{Van-Hau Pham} obtained his bachelor’s degree in computer science from the University of Natural Sciences of Hochiminh City in 1998. He pursued his master’s degree in Computer Science from the Institut de la Francophonie pour l’Informatique (IFI) in Vietnam from 2002 to 2004. Then he did his internship and worked as a full-time research engineer in France for 2 years. He then persuaded his Ph.D. thesis on network security under the direction of Professor Marc Dacier from 2005 to 2009. He is now a lecturer at the University of Information Technology, Vietnam National University Ho Chi Minh City (UIT-VNU-HCM), Hochiminh City, Vietnam. His main research interests include network security, system security, mobile security, and cloud computing.







\end{document}